\documentclass[twocolumn,showpacs,preprintnumbers,amsmath,amssymb]{revtex4-1}
\usepackage{graphicx}
\usepackage{dcolumn}
\usepackage{bm}
\usepackage[usenames,dvipsnames,svgnames,table]{xcolor}
\usepackage{SIunits}
\usepackage{textcomp}
\usepackage{tabularray}
\begin{document}
\title{Isotope substitution and polytype control for point defects identification: the case of the ultraviolet color center in hexagonal boron nitride}
\author{J. Plo,$^{1}$ A. Pershin,$^{2,8}$ S. Li,$^{2}$ T. Poirier,$^{3}$ E. Janzen,$^{3}$ H. Schutte,$^{3}$ M. Tian,$^{4}$ M. Wynn,$^{5}$ S. Bernard,$^{5}$ A. Rousseau,$^{1}$ A. Ibanez,$^{1}$ P. Valvin,$^{1}$ W. Desrat,$^{1}$ T. Michel,$^{1}$ V. Jacques,$^{1}$ B. Gil,$^{1}$ A. Kaminska,$^{6,7}$ N. Wan,$^{4}$ J. H. Edgar,$^{3}$ A. Gali,$^{2,8,9,\ast}$ G. Cassabois$^{1,10,\ast}$}
\affiliation{$^{1}$Laboratoire~Charles~Coulomb~UMR~5221~CNRS-Universit\'e~de~Montpellier, 34095 Montpellier, France\\
$^{2}$HUN-REN Wigner Research Centre for Physics, PO. Box 49, H-1525 Budapest, Hungary\\
$^{3}$Tim Taylor Department of Chemical Engineering, Kansas State University, Manhattan Kansas 66506, USA\\
$^{4}$Key laboratory of MEMS of Ministry of Education, School of Integrated Circuit, Southeast University, People's Republic of China\\
$^{5}$IRCER, CNRS, Univ. Limoges, 87068 Limoges, France\\
$^{6}$Institute of Physics, Polish Academy of Sciences, Al. Lotnikow 32/46, PL-02-668, Warsaw, Poland\\
$^{7}$Cardinal Stefan Wyszynski University, Faculty of Mathematics and Natural Sciences, School of Exact Sciences, Dewajtis 5, 01-815 Warsaw, Poland\\
$^{8}$Department of Atomic Physics, Institute of Physics, Budapest University of Technology and Economics,  M\H{u}egyetem rakpart 3., H-1111 Budapest, Hungary\\
$^{9}$MTA-WFK Lend\"ulet "Momentum" Semiconductor Nanostructures Research Group\\
$^{10}$Institut Universitaire de France, 75231 Paris, France}
\date{\today}
\begin{abstract}
Defects in crystals can have a transformative effect on the properties and functionalities of solid-state systems. Dopants in semiconductors are core components in electronic and optoelectronic devices. The control of single color centers is at the basis of advanced applications for quantum technologies. Unintentional defects can also be detrimental to the crystalline structure and hinder the development of novel materials. Whatever the research perspective, the identification of defects is a key but complicated, and often long-standing issue. Here, we present a general methodology to identify point defects by combining isotope substitution and polytype control, with a systematic comparison between experiments and first-principles calculations. We apply this methodology to hexagonal boron nitride (hBN) and its ubiquitous color center emitting in the ultraviolet spectral range. From isotopic purification of the host hBN matrix, a local vibrational mode of the defect is uncovered, and isotope-selective carbon doping proves that this mode belongs to a carbon-based center. Then, by varying the stacking sequence of the host hBN matrix, we unveil different optical responses to hydrostatic pressure for the non-equivalent configurations of this ultraviolet color center. We conclude that this defect is a carbon dimer in the honeycomb lattice of hBN. Our results show that tuning the stacking sequence in different polytypes of a given crystal provides unique fingerprints contributing to the identification of defects in 2D materials.
\end{abstract}
\maketitle
\section{Introduction}
Identifying defects is an important goal in condensed matter physics. Once the natures of extrinsic and intrinsic defects are determined, processes to improve the quality and the purity of crystals can be optimized. Conversely, the controlled incorporation of dopants either during or after the crystal growth are basic tools for fabricating electronic and optoelectronic devices. These routine strategies of the semiconductor industry are currently being revisited to develop quantum technologies, and reach the ultimate regime of single-defect creation and manipulation for advanced applications \cite{wolf}. Still, identifying defects is notoriously difficult, requiring the accumulation of complementary measurements. Even for silicon, where decades of research on defects of this flagship material were reviewed by Davies in Ref.\cite{davies} in the late eighties, only recently has the nature of some common color centers been elucidated \cite{Ggali,Wcenter}.

The advent of graphene and the flourishing of a large variety of 2D materials motivate a new wave of research on point defects. The novel properties and functionalities of atomically-thin crystals and their van der Waals heterostructures offer a new playground in solid-state physics, with plethora of unknown defects to identify, control and manipulate in the perspective of classical and quantum applications \cite{dresselhaus,crommie,turunen,lin}. Although the recent advances in \textit{ab initio} calculations are a key addition for screening possible candidates \cite{gali2023}, resolving the microscopic structure of defects remains challenging.

Here, we demonstrate a general methodology to identify point defects by combining isotope substitution and polytype control, with a systematic comparison between experiments and first-principles calculations. While isotope substitution is an established resource in semiconductor physics \cite{cardona}, tuning the stacking sequence in different polytypes of a given crystal makes use of a crucial degree of freedom, providing unique fingerprints to help identify defects in 2D materials. We apply this methodology to hexagonal boron nitride (hBN) and its ubiquitous color center emitting in the ultraviolet spectral range, to resolve the long-standing question of its nature.

hBN is a prototypical layered material \cite{geim}, with a simple honeycomb lattice of its basal plane where the orbitals of boron and nitrogen atoms are sp$^2$-hybridized with a strong in-plane covalent binding. In 2004, the synthesis of high-quality crystals triggered the rise of hBN in 2D materials research \cite{taniguchi2004}, but long before the pioneering work from NIMS-Japan, the existence of defects emitting at wavelengths $\sim$300 nm (energy $\sim$4 eV) was reported \cite{katzir}. Today, there is a large corpus of experimental and theoretical studies dealing with this ultraviolet color center, the so-called ``4 eV-defect'' \cite{pelini,taniguchi2007,katzir,kuzuba,museur,du,bourrellier,phuongPRL,tsushima,ChichibuJAP,koronski,vokhmintsev,vokhmintsev2,attaccalite,weston,korona,hamdi,mackoit,winter,li,rousseau,korona2023,kirchoff,li2024exceptionally}, however there is no consensus on its microscopic structure. On the one hand, Taniguchi \textit{et al.} \cite{taniguchi2007} showed that the emission of the 4 eV-defect is strongly enhanced in carbon-doped hBN, suggesting that this color center may contain carbon atoms. Recent \textit{ab initio} calculations \cite{weston,korona,mackoit,winter,li,kirchoff} proposed various carbon-based centers. On the other hand, Tsushima \textit{et al.} \cite{tsushima} found no correlation between the intensity of the photoluminescence (PL) signal of the 4 eV-defect and the carbon concentration in hBN, with possible carbon-free candidates such as the Stone-Wales defect which is predicted to precisely emit at $\sim$4 eV \cite{hamdi}.

\begin{center}
\begin{figure}
\includegraphics[width=0.45\textwidth]{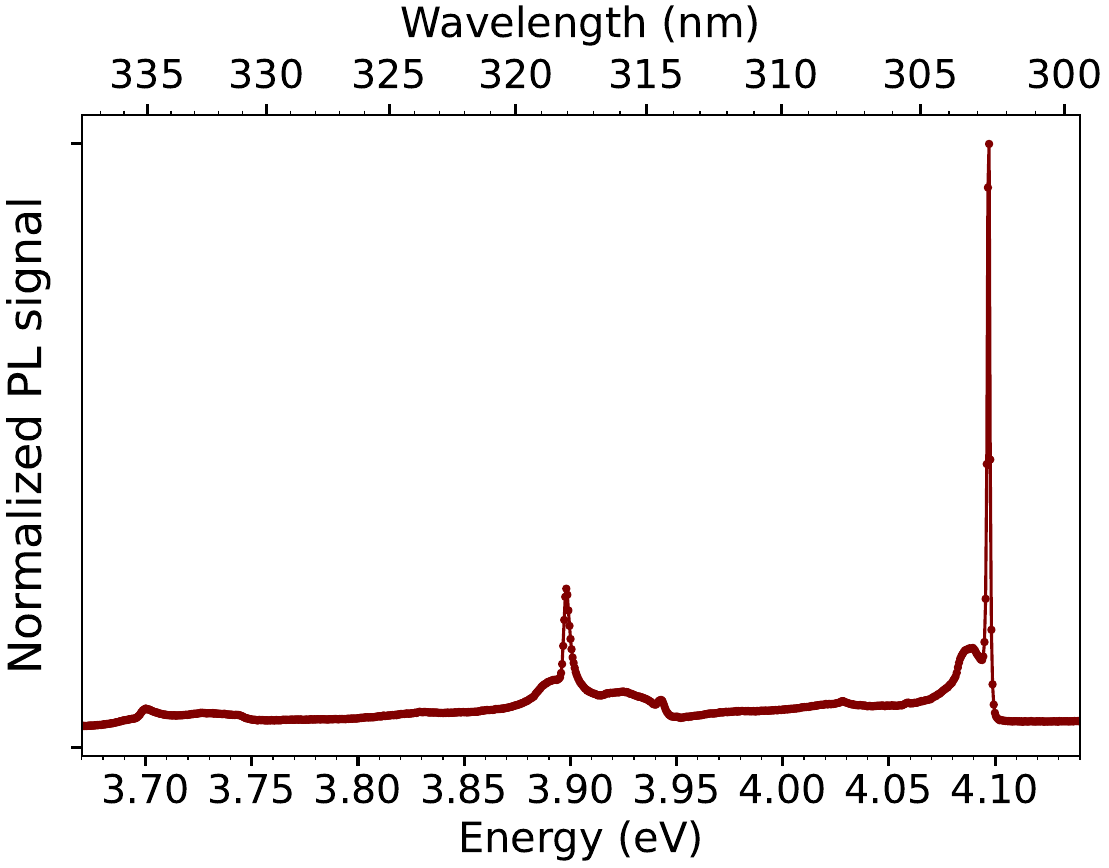}
\hfill
\includegraphics[width=0.4\textwidth]{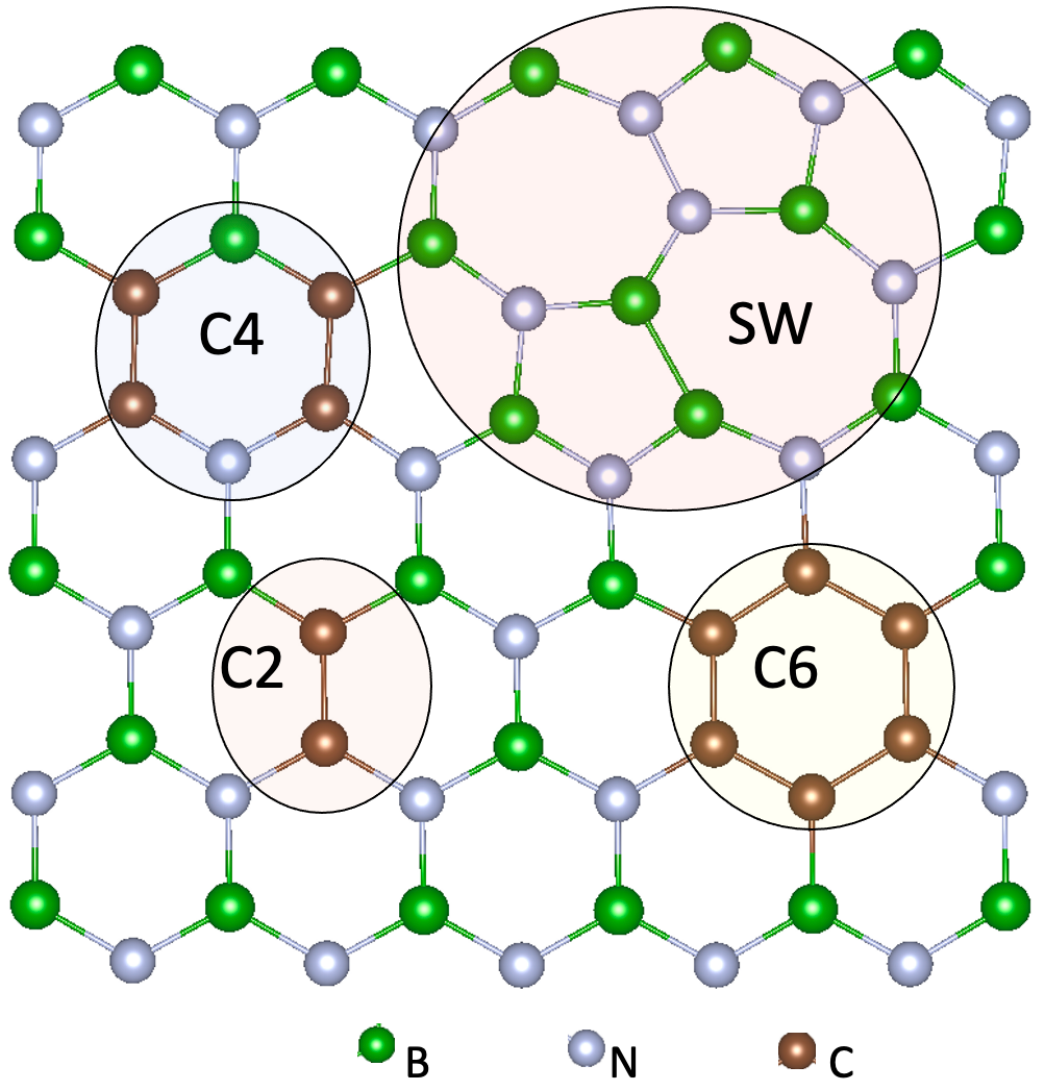}

\caption{(Top) Photoluminescence spectrum of the ultraviolet color center in hBN, the so-called 4 eV-defect, at low temperature (8 K). (Bottom) The four most probable structures identified in the literature are the carbon dimer (C2), the carbon tetramer (C4), the carbon 6-ring (C6) and the Stone-Wales defect (SW), which is an intrinsic defect consisting of two heptagons and two pentagons made out from four native hexagons.}
\label{fig0}
\end{figure}
\end{center}

The four most probable structures proposed so far for the 4 eV-defect are the carbon dimer (C2), the carbon tetramer (C4), the carbon 6-ring (C6) and the Stone-Wales defect (SW), as depicted in Fig.\ref{fig0} together with the typical PL spectrum detected in hBN in the corresponding spectral range. At low temperature, the emission spectrum of the 4 eV-defect is dominated by a sharp line at $\sim$4.09 eV, which is the zero-phonon line (ZPL) of this color center, i.e. the direct recombination without any net energy transfer with the phonon bath. At lower energy, the PL spectrum consists of a vibronic band composed of all phonon-assisted recombination channels in the defect.

By employing our methodology combining isotope substitution and polytype control, we present experimental and theoretical results that tackle the long-standing issue of the 4 eV-defect structure. We first implement isotopic purification of the host hBN matrix (Section \ref{sec:isotopehBN}), and we demonstrate that the phonon replica at $\sim$200 meV stems from a local vibrational mode (LVM) of the center, which is not consistent with the Stone-Wales structure. Then, we perform isotope-selective carbon doping of hBN and compare hBN doped with the $^{12}$C and $^{13}$C isotopes (Section \ref{sec:isotopeC}), and we resolve a 6 meV-isotopic shift of this LVM, unambiguously showing that the 4 eV-defect is a carbon-based center. Then, we vary the stacking sequence of the host hBN matrix and we perform pressure-dependent PL measurements of the 4 eV-defect in two hBN polytypes (Section \ref{sec:polytype}), either in the standard AA' stacking, or in the AB one, i.e. the Bernal form of hBN. From the different pressure-induced red-shifts of the ZPL in the two hBN polytypes, we conclude that the 4 eV-defect is a carbon dimer in the hBN honeycomb lattice.
\section{Isotopic purification of the host \MakeLowercase{h}BN matrix}
\label{sec:isotopehBN}
We first examine the impact of the isotopic purification of the host hBN matrix ($^{10}$B vs $^{11}$B, and $^{14}$N vs $^{15}$N) in samples fabricated in two facilities, following two variants of the same growth method.
\begin{center}
\begin{figure*}[ht]
\includegraphics[width=0.99\textwidth]{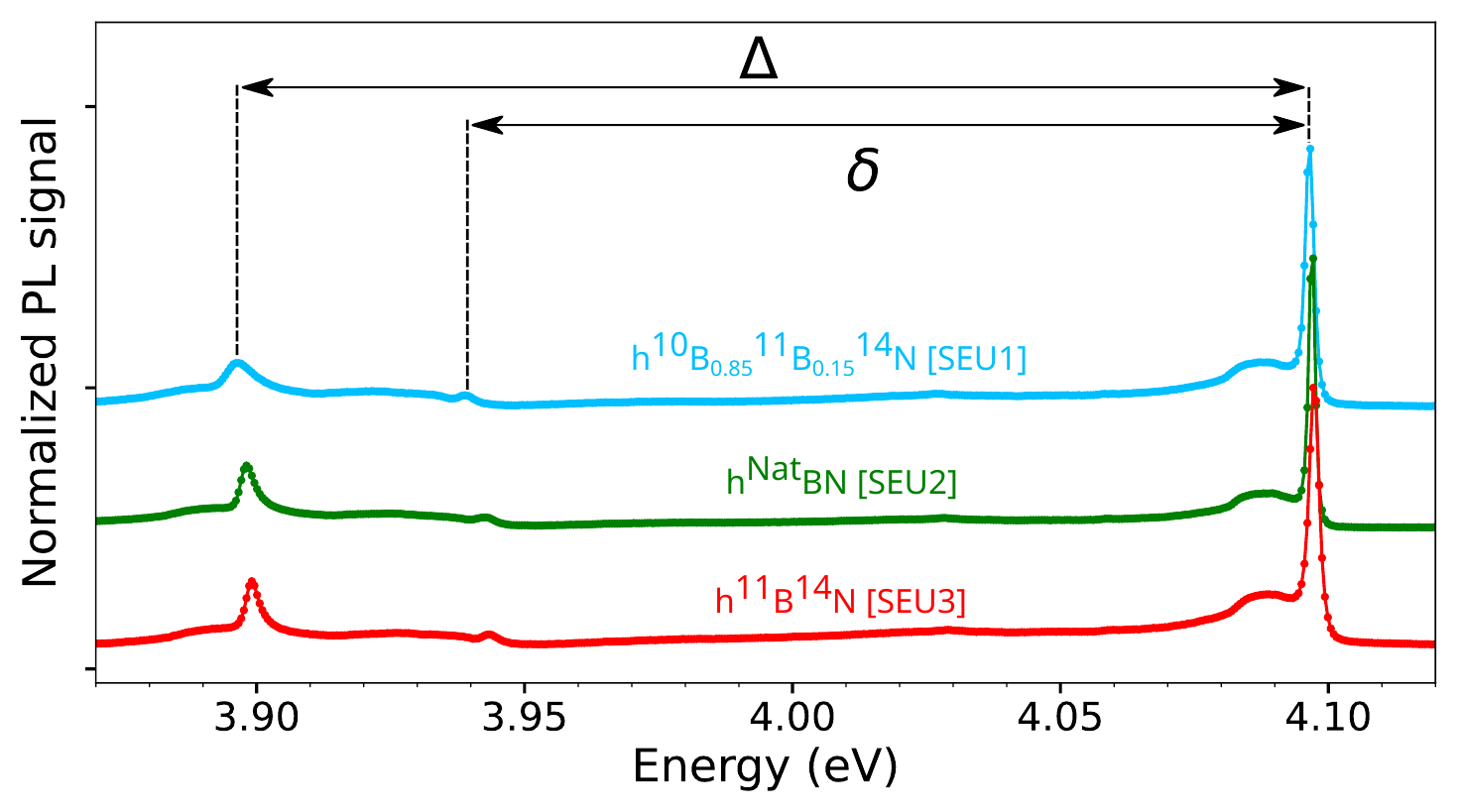}
\caption{PL spectra in the UV-B, at 8 K, recorded in isotopically-purified hBN samples from the SEU series (see detailed isotopic composition in Tab.\ref{multiprogram}). PL excitation energy = 6.32 eV. $\Delta$ is the energy splitting between the zero-phonon line at $\sim$4.1 eV and the phonon replica at $\sim$3.9 eV, and $\delta$ the splitting with the phonon replica at $\sim$3.945 eV.}
\label{fig1}
\end{figure*}
\end{center}
\begin{table}
  \centering

  \begin{tblr}{
    hline{1,2} = {2-Z}{solid},
    hline{3,6,11} = {solid},
    vline{1} = {3-Z}{solid},
    vline{2} = {solid},
    vline{4,Z} = {solid},  
    cells = {c},
    cell{1}{2} = {c=2}{c},
    cell{1}{4} = {c=2}{c}, 
  }

        & Boron    &  & Nitrogen &   \\
        & $^{10}$B & $^{11}$B & $^{14}$N & $^{15}$N \\
    SEU1  & 85 & 15  & 99.6  & 0.4     \\
    SEU2  & 19.9 & 80.1  & 99.6  & 0.4     \\
    SEU3  & $<$1 & $>$99  & 99.6  & 0.4     \\
    KSU1  & 99.2 & 0.8  & 99.6  & 0.4     \\
    KSU2  & 99.2 & 0.8  & $<$1  & $>$99     \\
    KSU3  & 19.9 & 80.1  & 99.6  & 0.4     \\
    KSU4  & 0.6 & 99.4  & 99.6  & 0.4     \\
    KSU5  & 0.6 & 99.4  & $<$1  & $>$99     \\
\end{tblr}
\caption{Isotopic content of hBN crystals fabricated at SouthEast University (SEU series), and Kansas State University (KSU series). SEU2 and KSU3 are h$^{\rm{Nat}}$BN crystals grown with naturally abundant boron and nitrogen isotopes. All samples in the SEU and KSU series correspond to hBN in the AA' stacking.}
\label{multiprogram}
\end{table}
\subsection{Samples}
Samples from SouthEast University (SEU series) were grown by using the metal flux method [see Appendix~\ref{sec:BN_SEU}]. For boron isotope controlled single crystal growth, powders of $^{10}$B (85 at\%) or $^{11}$B (99 at\%) were used as source materials for the synthesis of $^{10}$B-enriched hBN or h$^{11}$B$^{14}$N crystals, respectively. Carbon powder (with naturally abundant isotopes, i.e. 99\% $^{12}$C and 1\% $^{13}$C) was included in all growths to promote the hBN single crystal formation \cite{wan}. Following the characterization of hBN crystals with a controlled composition in $^{10}$B and $^{11}$B and naturally abundant nitrogen \cite{vuong,cusco}, the boron isotope content of SEU1 and SEU3 was checked by Raman and UV-C PL spectroscopy measurements.

The hBN crystals from Kansas State University (KSU series) were synthesized through the metal flux growth method described in Ref.\cite{janzen}. The source materials are (i) boron powders isotopically enriched with either $^{10}$B (99.2\%) or $^{11}$B (99.4\%), and (ii) a nitrogen gas either featuring a natural $^{14}$N content, or enriched with the $^{15}$N ($>$99\%) isotope. Raman and UV-C PL spectroscopy have recently confirmed that such a growth method provides high-quality hBN crystals in the four different configurations of the stable boron and nitrogen isotopes, namely h$^{10}$B$^{14}$N, h$^{10}$B$^{15}$N, h$^{11}$B$^{14}$N, and h$^{11}$B$^{15}$N \cite{janzen}.

The isotope compositions of the eight samples studied in Section \ref{sec:isotopehBN} are summarized in Tab.\ref{multiprogram}. SEU2 and KSU3 are reference samples grown with naturally abundant boron and nitrogen isotopes. All samples in the SEU and KSU series correspond to hBN in the standard AA' stacking.
\begin{center}
\begin{figure}[ht]
\includegraphics[width=0.49\textwidth]{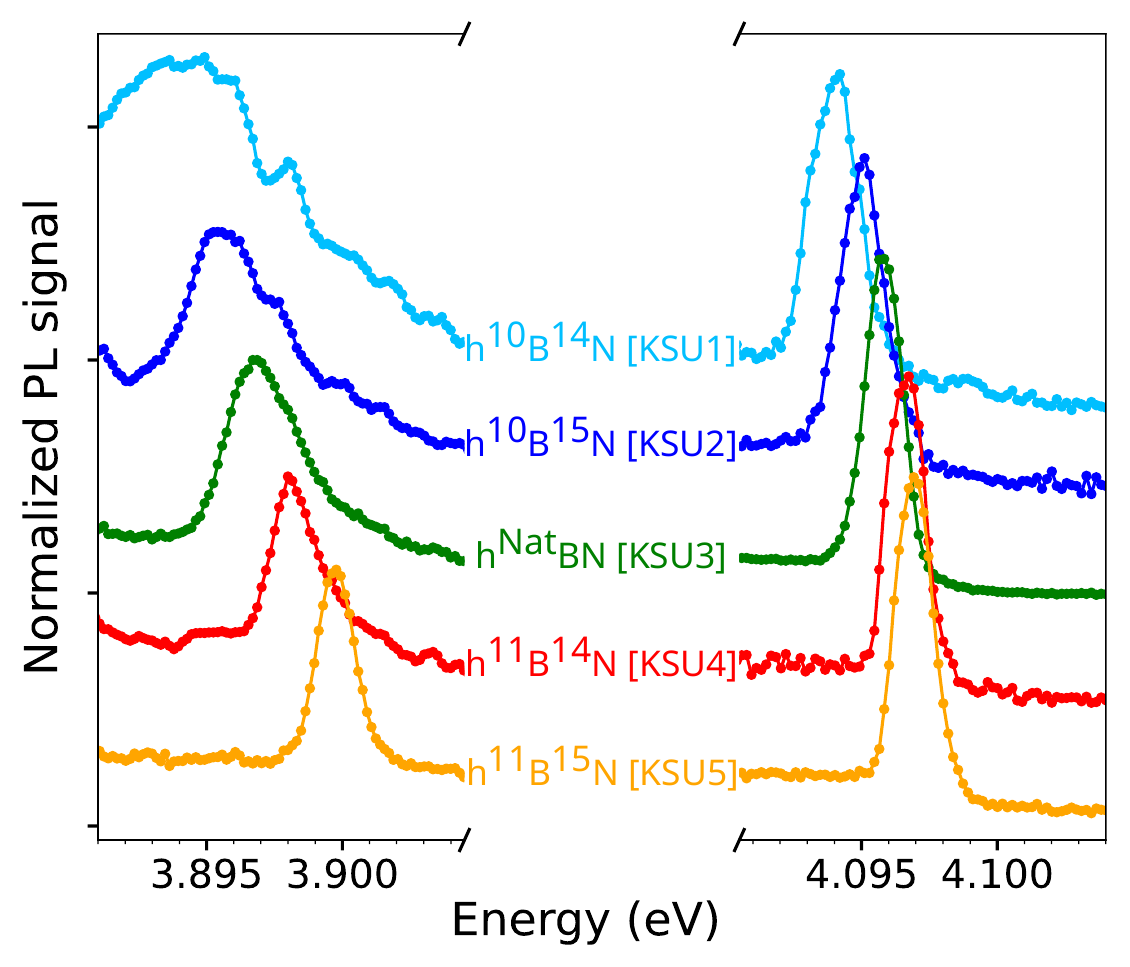}
\caption{PL spectra in the UV-B, at 8 K, recorded in isotopically-purified hBN samples from the KSU series (see detailed isotopic composition in Tab.\ref{multiprogram}), zoomed around the phonon replica at $\sim$3.9 eV and the zero-phonon line at $\sim$4.1 eV. PL excitation energy = 6.32 eV.}
\label{fig2}
\end{figure}
\end{center}
\begin{center}
\begin{figure*}[ht]
\includegraphics[width=1\textwidth]{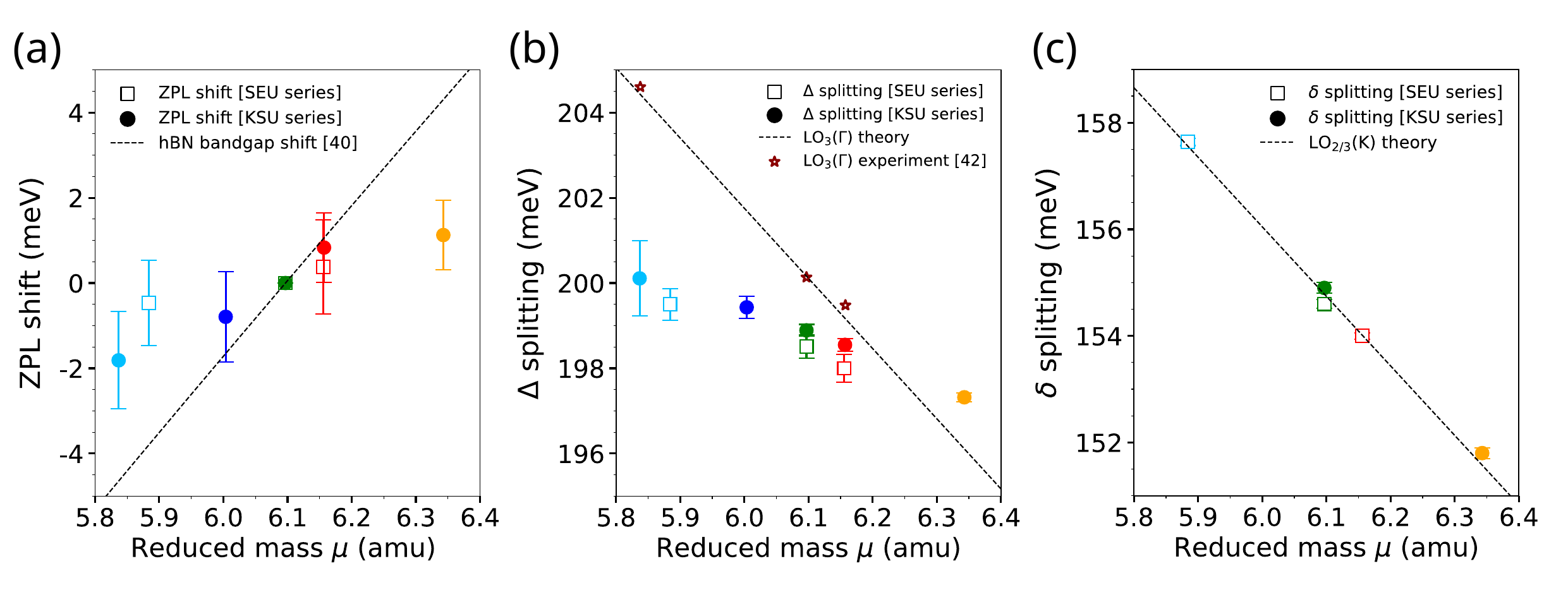}
\caption{(a) Isotopic shift of the zero-phonon line at $\sim$4.1 eV with respect to h$^{\rm{Nat}}$BN crystals, grown with naturally abundant boron and nitrogen isotopes, as a function of the reduced mass $\mu$ in units of the atomic mass (amu). The dashed line corresponds to the experimental isotopic shift of the hBN bandgap at $\sim$5.95 eV, from Ref.\cite{janzen}. (b) Isotopic shift of the energy splitting $\Delta$, between the zero-phonon line at $\sim$4.1 eV and the phonon replica at $\sim$3.9 eV (as defined in Fig.\ref{fig1}), as a function of the reduced mass $\mu$. The dashed line corresponds to the calculated isotopic shift of the LO$_3$ optical phonon at the center of the Brillouin zone, and stars indicate the experimental values, from Ref.\cite{giles}. (c) Isotopic shift of the energy splitting $\delta$, between the zero-phonon line at $\sim$4.1 eV and the phonon replica at $\sim$3.945 eV (as defined in Fig.\ref{fig1}), as a function of the reduced mass $\mu$. The dashed line corresponds to the calculated isotopic shift of the LO$_{2/3}$(K) optical phonon at the K point of the Brillouin zone. In all panels, experimental data are displayed as empty squares for the SEU series, and full circles for the KSU one, with the same color code as Fig.\ref{fig1}\&\ref{fig2}.}
\label{fig3}
\end{figure*}
\end{center}
\begin{center}
\begin{figure*}[ht]
\includegraphics[width=0.99\textwidth]{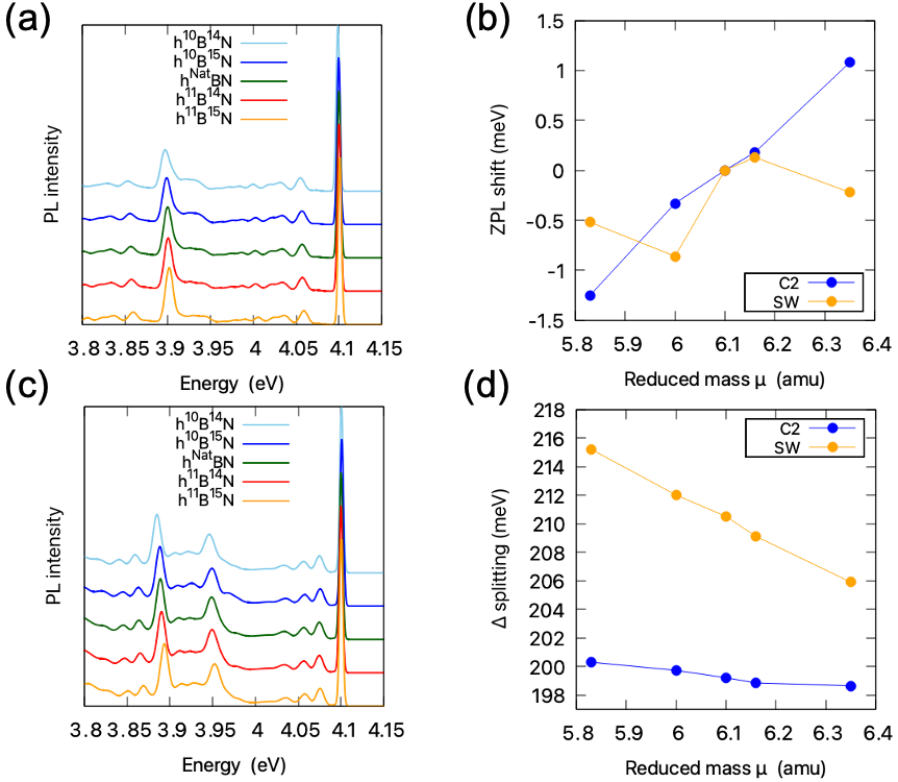}
\caption{Phonon sidebands, calculated for (a) C2 and (c) SW defects in hBN of different isotope composition. (b) Calculated isotopic shift of the zero-phonon line at $\sim$4.1 eV for C2 and SW defects. (d) Calculated isotopic shift of the energy splitting $\Delta$, between the zero-phonon line at $\sim$4.1 eV and the phonon replica (as defined in Fig.\ref{fig1}) for C2 and SW defects.}
\label{fig_t1}
\end{figure*}
\end{center}
\subsection{Experimental results}
Fig.\ref{fig1} displays the PL spectra measured in hBN crystals of the SEU series, with the experimental setup described in Appendix~\ref{sec:PL_MPL}. The vibronic band of the 4 eV-defect involves low-energy acoustic phonons at energies $\sim$4.08-4.09 eV close to the ZPL, and also phonons spanning the full bandstructure as witnessed by the phonon-photon mapping in the vibronic band in this color center \cite{phuongPRL}. We define $\delta$ as the energy splitting between the ZPL at $\sim$4.1 eV and the phonon replica at $\sim$3.945 eV, and $\Delta$ as the one with the replica at $\sim$3.9 eV [Fig.\ref{fig1}]. The phonon replica at $\sim$3.9 eV has so far been assigned to the recombination assisted by the emission of a LO$_3$($\Gamma$) optical phonon at the center of the Brillouin zone \cite{phuongPRL,cusco2016}, because the energy splitting $\Delta$ with the ZPL matches the LO$_3$($\Gamma$) phonon energy. We will see below that our measurements in isotopically-purified hBN crystals rebut this initial interpretation. In what follows, we focus on the isotopic shifts of the ZPL energy, and of the $\Delta$ and $\delta$ splittings.

In the full-scale display of the SEU series in Fig.\ref{fig1}, the PL spectra of the 4 eV-defect in isotopically-purified hBN crystals appear very similar. There are however some discernible differences for the narrowest lines. First, the ZPL and the phonon replica at $\sim$3.9 eV have widths that change from one sample to another. Because the latter belongs to the satellite-vibronic band of the former, the broader the ZPL, the larger the width of the phonon replica, as seen in the temperature-dependent measurements of Ref.\cite{phuongPRL}. Here, the ZPL width is 2, 1.6 and 2.2 meV from SEU1 to SEU3, while the phonon replica at $\sim$3.9 eV monotonously narrows with a width of 5.6, 3.2 and 3 meV, respectively. We thus conclude that the width variations of the phonon replica are not related to the inhomogeneous broadening of the ZPL, pointing to an intrinsic origin of this effect. Second, the ZPL and phonon replica shift within the SEU series [Fig.\ref{fig1}]. Albeit hardly observable in Fig.\ref{fig1}, there is a red-shift of the lines for light isotopes. 

This systematic effect of the isotopic purification is better resolved for the KSU series in Fig.\ref{fig2}. In Fig.\ref{fig2}, we present the PL spectra of the KSU series [Tab.\ref{multiprogram}], with the four configurations of the stable boron and nitrogen isotopes, h$^{10}$B$^{14}$N, h$^{10}$B$^{15}$N, h$^{11}$B$^{14}$N, h$^{11}$B$^{15}$N, together with a reference h$^{\rm{Nat}}$BN sample (KSU3). In contrast to the SEU series, no carbon was intentionally added during the crystal growth, leading to a weaker PL signal at $\sim$4 eV, in agreement with previous reports \cite{taniguchi2007}. The 4 eV-defect emission is dim in the KSU series because of the high-purity of the boron and nitrogen precursors employed for the growth of isotopically-purified hBN. This point is of great importance in the prospect of the isotope-selective carbon doping, as later discussed in Section \ref{sec:isotopeC}. Because of the sparse emission, recording the PL spectrum of the 4 eV-defect required the use of micro-PL measurements [see Appendix~\ref{sec:PL_MPL}], to locate the areas of highest defects density in crystals of the KSU series, hence the reduced signal-to-noise ratio in Fig.\ref{fig2} compared to Fig.\ref{fig1}. To better observe the isotopic shifts of the ZPL and phonon replica, we use a split x-axis representation in Fig.\ref{fig2}, where we zoom in around two intervals of the same 13 meV-extension, centered at 3.8975 and 4.0975 eV. There, both the ZPL and the phonon replica systematically red-shift as a function of the isotopic masses. Interestingly, the ZPL energy only changes by $\sim$3 meV in contrast to $\sim$5 meV for the phonon replica, thus revealing the dependence of the energy splitting $\Delta$ [as defined in Fig.\ref{fig1}] with the isotopic composition. Similar to the SEU series, the phonon replica is the broadest for the lightest isotopes (h$^{10}$B$^{14}$N, KSU1), further supporting the intrinsic origin of this effect.

In Fig.\ref{fig3}a, we merge the data of the SEU and KSU series for the isotopic shift of the ZPL. For each series, we subtract the ZPL energy measured in the reference h$^{\rm{Nat}}$BN sample (4.0988 eV for SEU2 and 4.0982 eV for KSU3) to the value found in the different isotopically-purified hBN crystals. We follow this procedure because the two series were measured in different runs. The 0.6 meV-difference between SEU2 and KSU3 lies within our absolute calibration accuracy, thus showing the repeatability of the 4 eV-defect formation in hBN crystals, not only grown at different growth facilities, but also with different procedures where the carbon quantity significantly varies.

Following Ref.\cite{vuong}, we plot the isotopic shift of the ZPL as a function of the reduced mass $\mu$, expressed in units of the atomic mass as:
\begin{equation}
\frac{1}{\mu}=\frac{1}{10x+11(1-x)}+\frac{1}{14y+15(1-y)}
\end{equation}
where $x$ and $y$ are the $^{10}$B and $^{14}$N percentages of the boron and nitrogen isotopes, respectively. The isotopic shifts measured for the SEU (empty squares) and KSU (full circles) hBN crystals assemble in a self-consistent way in Fig.\ref{fig3}a, demonstrating the universal impact of isotopic purification of the host hBN matrix on the ZPL of the 4 eV-defect. We compare these data to the isotopic shift of the hBN bandgap [dashed line in Fig.\ref{fig3}a], which is due, at low temperatures, to the zero-point vibrations mediated by the electron-phonon coupling \cite{cardona}. It was first estimated in hBN in Ref.\cite{vuong} and more recently in Ref.\cite{janzen} with the extension of isotopic purification to $^{15}$N. The $\sim$3 meV-shift of the 4 eV-defect ZPL appears to be three times smaller than the hBN bandgap shift, over the investigated range of isotopic masses. Such a discrepancy is the signature of the electronic localization in a deep level, interpreted below with our first-principles calculations.

Fig.\ref{fig3}b displays the merged data of the SEU and KSU series for the energy splitting $\Delta$, as defined in Fig.\ref{fig1}, as a function of the reduced mass $\mu$. Similarly to the ZPL isotopic shift in Fig.\ref{fig3}a, the congruent measurements in samples fabricated in two different growth facilities reveal the impact of the isotope control in hBN on the energy splitting $\Delta$. Because $\Delta$ corresponds to a characteristic vibrational energy in phonon-assisted recombination processes, its variations with the isotopic masses is straigthforward, in contrast to the ZPL isotopic shift which stems from more subtle effects related to the electron-phonon coupling and the zero-point vibrations. Specifically, $\Delta$ decreases with the reduced mass $\mu$ [Fig.\ref{fig3}b] because the frequency of an harmonic oscillator is inversely propotional to the square root of its mass, as observed by Raman spectroscopy in hBN in Ref.\cite{vuong,janzen}. Because $\Delta$ is of the order of 200 meV, we compare its isotopic shift to the one of the LO$_3$ optical phonon at the center of the Brillouin zone. In Fig.\ref{fig3}b, calculations of the LO$_3$($\Gamma$) energy within the QuantumEspresso package are displayed in dashed line, while the experimental values measured in Ref.\cite{giles} correspond to the stars. Strikingly, in h$^{\rm{Nat}}$BN ($\mu$=6.1), the energy splitting $\Delta$ almost coincides with the LO$_3$($\Gamma$) energy. This is why the phonon replica at $\sim$3.9 eV was initially interpreted as involving the emission of one LO$_3$($\Gamma$) phonon \cite{phuongPRL}. However, varying the isotopic composition leads to distinct variations, thus unveiling that $\Delta$ is not the energy of one LO$_3$($\Gamma$) phonon, and that the phonon replica at $\sim$3.9 eV stems from a LVM of the 4 eV-defect.

It is instructive to compare the $\delta$ splitting, between the ZPL and the phonon replica at $\sim$3.945 eV, as defined in Fig.\ref{fig1}. In that case, the variations of $\delta$ with the reduced mass $\mu$ nicely follow the calculated isotopic shift of the LO$_{2/3}$(K) phonon, i.e. the longitudinal optical mode of the quasi-degenerate 2 and 3 branches at the K point of the Brillouin zone [dashed line, Fig.\ref{fig3}c]. This Bloch phonon mode of the hBN matrix gives rise to a large peak at $\sim$155 meV in the phonon density states, hence the phonon replica at $\sim$3.945 eV can be attributed to recombination assisted by the emission of a LO$_{2/3}$(K) optical phonon \cite{phuongPRL}. Our experiments in isotopically-purified hBN crystals confirm this interpretation, and we will make use of this complementary fingerprint when analysing the isotope-selective carbon doping in Section \ref{sec:isotopeC}.


\subsection{\textit{Ab initio} calculations}
After presenting the first part of our experimental findings, we now turn our attention to a theoretical identification of the defect structure responsible for the isotopic response observed in the 4.1 eV-PL signal. 
We apply highly convergent supercell plane wave density functional theory calculations to calculate the PL signals of selected defects [see Appendix~\ref{sec:DFT_addit} for details].
In Fig.\ref{fig0}, we illustrate the structures of four defects: the Stone-Wales (SW) defect and three different complexes based on carbon impurities, all of which have previously been suggested as potential emitters of the 4.1 eV signal. Initially, due to the strong similarities in the PL spectra of carbon-based defects in an isotopically-purified hBN matrix, we focus on the carbon dimer (C2) for our spectral simulations, as a representative example of a defect made of an even number of carbon impurities in hBN. 

In Fig.\ref{fig_t1}, we show the calculated modifications to the PL spectrum of these defects with variations in the isotope composition of the host material. Our simulations reveal that the PL signal of the C2 defect closely matches our experimental findings, displaying features such as a prominent phonon replica $\sim$200 meV below the ZPL, and a shoulder at $\sim$170 meV. In contrast, the SW defect exhibits two groups of active modes at around 210 and 150~meV. The 210 meV mode is localised at the boron and nitrogen atoms, connecting two pentagons [see Fig.\ref{fig_ts1} in Appendix~\ref{sec:SI_DFT}]. The three active modes at $\sim$150~meV are hybridised with the bulk modes, with a particular impact on the atoms within the nitrogen-rich pentagon.
For both types of defects, the phonon sidebands are primarily influenced by stretching modes, as depicted in Figs.\ref{fig_ts1} and \ref{fig_ts1a} in Appendix~\ref{sec:SI_DFT}. Additionally, the partial Huang-Rhys factors for other active modes, which account for additional peaks in our simulated vibronic bands, are also presented in Figs.\ref{fig_ts1} and \ref{fig_ts1a}.

Fig.\ref{fig_t1}b further displays the calculated isotopic shift of the ZPL for the C2 defect, which increases systematically with the reduced mass $\mu$ - consistently with our experimental observations. Conversely, a less clear trend is observed for the SW defect, which  is likely due to an imbalanced participation of nitrogen and boron atoms to the optical transition \cite{hamdi}. 
Most strikingly, the vibrational energy of the stretching mode in the SW defect, involving both boron and nitrogen atoms, exhibits a pronounced decrease of $\sim$9 meV with changes in the isotope composition of hBN, as shown in Fig.\ref{fig_t1}d. 
This shift is comparable to the one of a Bloch mode of the hBN matrix, such as the LO$_3$($\Gamma$) phonon [Fig.\ref{fig3}b]. In contrast, the energy of the carbon-localized mode in C2 only marginally decreases by less than 2 meV. A direct comparison between the calculated $\Delta$ splittings and our experimental data is provided in Fig.\ref{fig_ts2} in Appendix~\ref{sec:SI_DFT}, where the three carbon-based defects (C2, C4 and C6) are considered. Notably, the computed $\Delta$ splittings for the C2 defect closely align with our experimental results, whereas for the carbon tetramer (C4), they are systematically overestimated by 4 meV, and underestimated by 2 meV for the carbon 6-ring (C6). Therefore, given the quantitative agreement between our experimental and theoretical results, we infer that the C2 defect is so far the most likely candidate for the 4.1 eV-emission.
\begin{center}
\begin{figure*}[ht]
\includegraphics[width=0.99\textwidth]{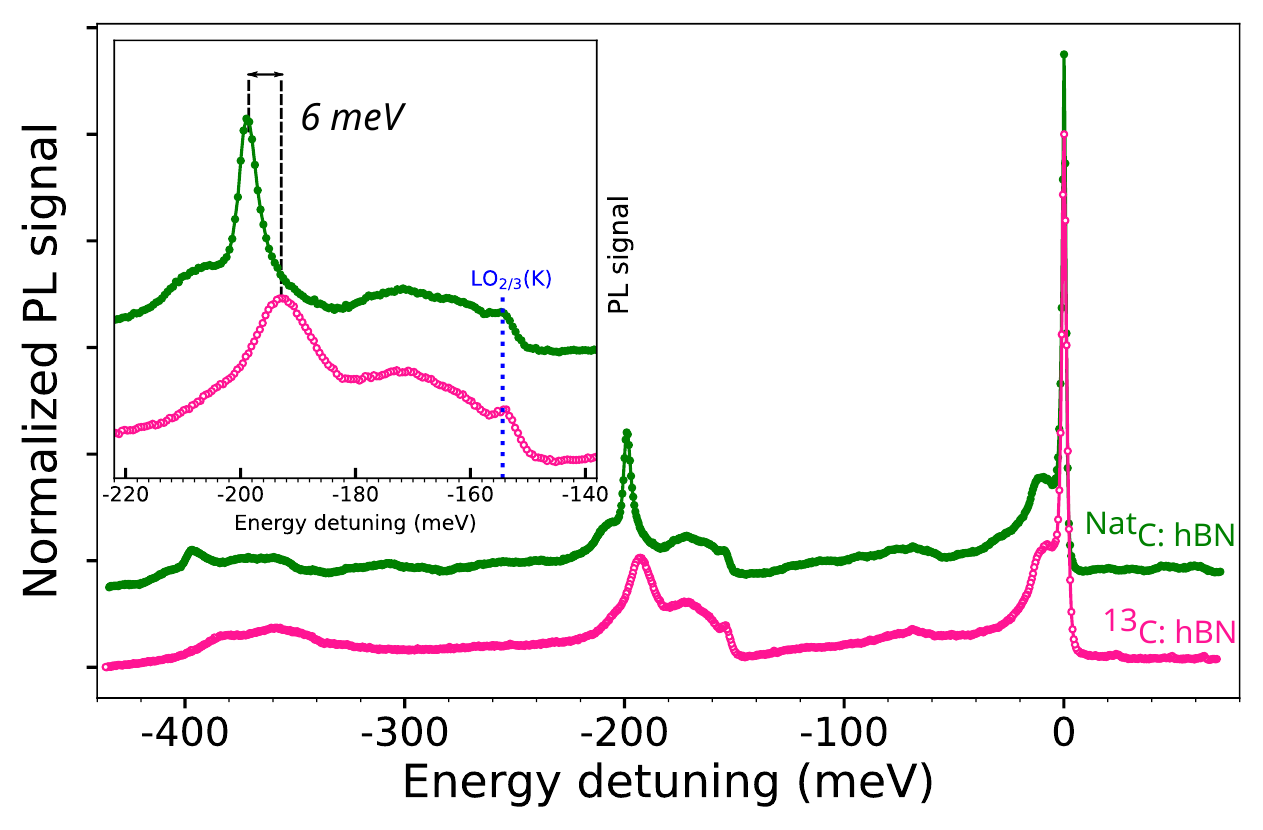}
\caption{Impact of the isotopic substitution ($^{\rm{Nat}}$C vs $^{13}$C) on the PL spectrum of the 4 eV-defect, plotted as a function of the energy detuning with the ZPL energy, in h$^{\rm{Nat}}$BN crystals grown with naturally abundant boron and nitrogen isotopes. PL excitation energy = 6.32 eV, sample temperature = 8 K. Inset: zoom around the local vibrational mode at $\sim$200 meV.}
\label{fig6}
\end{figure*}
\end{center}
\begin{center}
\begin{figure*}[ht]
\includegraphics[width=0.99\textwidth]{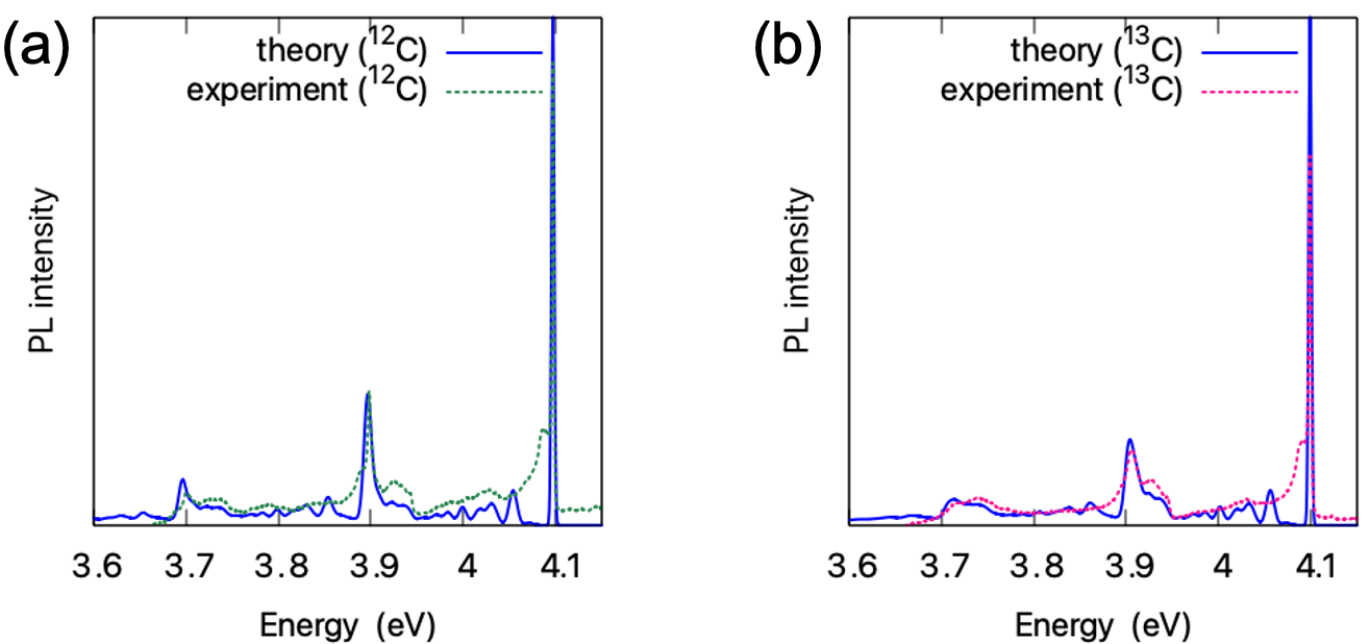}
\caption{Calculated PL spectrum of the carbon dimer (C2) for either $^{12}$C (a) or $^{13}$C (b) impurities. First-principles simulations (solid lines) are compared to experimental results of Fig.\ref{fig6} (dotted lines).}
\label{fig_t2}
\end{figure*}
\end{center}
\section{Isotope-selective carbon doping}
\label{sec:isotopeC}
Having identified the phonon replica at $\sim$3.9 eV as stemming from a LVM, and having ruled out the Stone-Wales center [Fig.\ref{fig0}] as a compatible structure for the isotopic shift of this LVM, we examine in this section the impact of isotope-selective carbon doping for further testing and screening the carbon-based defects represented in Fig.\ref{fig0}, among which the carbon dimer appears as a suitable candidate at this stage.
\subsection{Samples}
hBN crystals were grown in KSU by precipitating from a solution formed by dissolving either $^{10}$B or $^{11}$B enriched boron in molten iron. The solution was formed by heating to 1550\degreecelsius, holding at this temperature to homogenize the solution, then slowly cooling to 1500\degreecelsius\, at 1 \degreecelsius/h (compared to 4 \degreecelsius/h in previous studies \cite{pelini}). Nitrogen (250 sccm) and forming gas: argon (95\%) hydrogen (5\%) (50 sccm) continuously flowed over the solution during the process.

For isotope-selective carbon doping, graphite powder was added to the starting materials, either with the natural distribution of isotopes (approximately 99\% $^{12}$C and 1\% $^{13}$C) or enriched in $^{13}$C (99\% with 1\% $^{12}$C). For $^{\rm{Nat}}$C-doping, graphitic carbon (99.995\% purity) was loaded at a 0.44 mass ratio to that of the boron. The total composition was 96.65 mass percent iron, 2.33 mass percent natural boron, and 1.02 mass percent carbon. For isotopically enriched $^{13}$C-doping, isotopic carbon (97.1\% chemical, 99\% isotope purity) was used in lieu of the natural carbon.

The carbon doping process has been refined to be more reproducible since what was first reported in Ref.\cite{pelini}. In that work, the boron source was exclusively boron nitride; in this work, elemental boron powder (99\%) was used. Furthermore, in this study, carbon monoxide was not used; only forming gas to getter oxygen. Additionally, the solvent was pure iron (99.99\% purity) contrary to the mixtures reported in Ref.\cite{pelini}. Overall, the residual $^{\rm{Nat}}$C content relative to the boron in each sample is significantly lower than those reported by Pelini \textit{et al} \cite{pelini}. This point was already commented in Section \ref{sec:isotopehBN} when presenting the measurements by micro-PL in the KSU series [Fig.\ref{fig2}]. This is a major difference compared to previous studies where unintentional $^{\rm{Nat}}$C contamination presumably hindered the impact of $^{13}$C-doping \cite{pelini}, and it is the key here for revealing the fingerprints of isotope-selective carbon doping in hBN.
\subsection{Experimental results}
The PL spectra recorded in $^{\rm{Nat}}$C:hBN and $^{13}$C:hBN crystals are plotted in Fig.\ref{fig6} as a function of the detuning with the ZPL energy, in order to eliminate any isotopic shift of the ZPL and solely focus on the influence of the carbon isotopes on the vibronic band spectrum. When replacing $^{\rm{Nat}}$C (i.e. $^{12}$C at 99\%) with $^{13}$C, modifications occur in two ranges of energy detuning: (i) at $\sim$200 meV, which corresponds to the LVM energy identified in Section \ref{sec:isotopehBN}, and (ii) at $\sim$400 meV where phonon-assisted recombination involves two-phonon overtones of the 200 meV-LVM. Zooming in around 200 meV-detunings resolves a 6 meV-blue-shift of the phonon replica in $^{13}$C:hBN and also a line-broadening [Fig.\ref{fig6}, inset]. Because the ZPL width hardly changes from 2.2 to 2.8 meV in $^{\rm{Nat}}$C:hBN and $^{13}$C:hBN, the latter effect does not come from an increased inhomogeneous broadening in $^{13}$C:hBN crystals, thus pointing for an intrinsic origin, similarly to our previous results in Section \ref{sec:isotopehBN} dealing with the impact of the isotopic purification of the host matrix.

The blue-shift of the phonon replica at $\sim$3.9 eV implies a reduced energy of the LVM, which is consistent with a carbon-based color center where the heavier mass of the $^{13}$C isotope with respect to $^{\rm{Nat}}$C results in vibrational modes of lower energy \cite{hopfield,daviesSi}. Importantly, the phonon-assisted recombination line at $\sim$3.945 eV, corresponding to an energy detuning $\delta\sim155$ meV, is not affected by the $^{13}$C vs $^{\rm{Nat}}$C selective carbon doping [Fig.\ref{fig6}, inset]. This effect is perfectly in line with our findings of Section \ref{sec:isotopehBN} where we concluded that the $\sim$3.945 eV line originates from the recombination assisted by the emission of a Bloch phonon mode of the hBN matrix. For moderate carbon concentrations, the phonon bandstructure of hBN is unperturbed by the presence of carbon impurities, so that any phonon-assisted recombination line involving a delocalized vibrational mode will not be influenced by isotope-selective carbon doping.

From the reduced $\Delta$ splitting in $^{13}$C:hBN on the one hand, and the constant $\delta$ value on the other hand, we unambiguously conclude that the 4 eV-defect is a carbon-based defect in hBN. Our results thus resolve the long-standing controversy on the interplay between carbon and this ultraviolet color center in hBN, carbon being interpreted either as a constituent of the 4 eV-defect, or only as a vector of its formation, or even as a simple bystander \cite{pelini,taniguchi2007,katzir,kuzuba,museur,du,bourrellier,phuongPRL,tsushima,ChichibuJAP,koronski,vokhmintsev,vokhmintsev2,attaccalite,weston,korona,hamdi,mackoit,winter,li,kirchoff,rousseau,korona2023}.
\subsection{\textit{Ab initio} calculations}
In order to further quantitatively interpret our experimental results, we conducted first-principles calculations [see Appendix~\ref{sec:DFT_addit}] to examine the impact of carbon isotopes on the optoelectronic properties of the three carbon-based candidate structures [Fig.\ref{fig0}]. First, we focus on analyzing the response of the carbon dimer (C2), which was identified as the most probable defect structure in section II.C. Fig.\ref{fig_t2} illustrates a direct comparison between our theoretical and experimental spectra. Our calculations for both $^{12}$C2 and $^{13}$C2 defects exhibit a fair agreement with our experimental data, particularly capturing the characteristic broadening of the first and second phonon replica attributed to the $^{13}$C isotope. The calculated variation of the $\Delta$ splitting was determined to be 6~meV, which is the exact experimental value [Fig.\ref{fig6}]. 

Notably, our analysis revealed a slight underestimation of the computed shift by $\sim$2~meV, when considering only the energy of the primary stretching mode. This effect can be attributed to two additional modes at around 190~meV that are partially localized on the carbon atoms [see Fig.~\ref{fig_ts1a}a in Appendix~\ref{sec:SI_DFT}], shaping the vibration peak in the phonon sideband of PL. The combined influence of these three modes explains the change in the isotope-dependent broadening of the vibration peak at around 200~meV, caused by the carbon isotopes. Interestingly, the relative contribution of the two additional modes at $\sim$190~meV is also influenced by the nitrogen and boron isotopes, with a significant increase observed for the lighter atoms [see Fig.~\ref{fig_ts1a}b in Appendix~\ref{sec:SI_DFT}]. This explains the systematic broadening of the band at $\sim$3.9~eV for the lighter boron and nitrogen isotopes, as observed in our experiments presented in Section~\ref{sec:isotopehBN}. These findings emphasize the importance of conducting a comprehensive spectrum simulation to accurately capture the isotopic effects for defects in hBN.

When examining the two other carbon defects, we observe a poorer agreement with our experimental data [see Appendix~\ref{sec:SI_DFT}, Fig.\ref{fig_ts3}]. Specifically, the carbon tetramer (C4) also displays a 6 meV-isotopic shift of the $\Delta$ splitting, but at a higher vibrational energy, particularly evident at emission energies $\sim$3.7 eV where phonon-assisted recombination involves two-phonon overtones of the 200 meV-LVM. The carbon 6-ring (C6) exhibits a larger isotopic shift of 9 meV for the $\Delta$ splitting, along with an overall greater mismatch between the energies of the active vibrational modes. These discrepancies further support our conclusion that the carbon dimer is the likely defect responsible for the observed 4.1 eV-signal in hBN.
\begin{center}
\begin{figure*}[ht]
\includegraphics[width=0.99\textwidth]{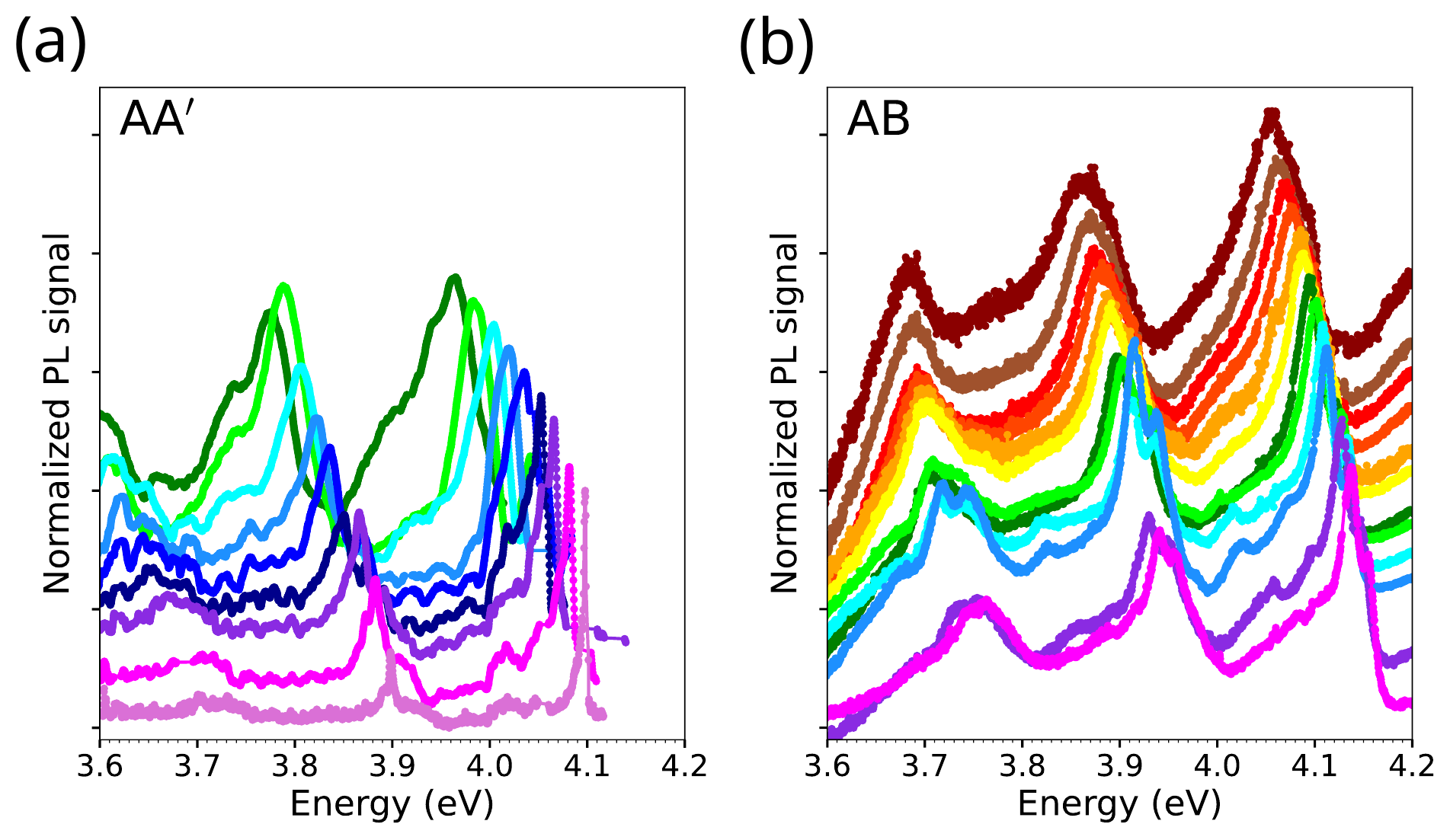}
\caption{Impact of the stacking sequence (AA' vs AB) of the host hBN matrix on the pressure-dependent PL response of the 4 eV-defect in h$^{\rm{Nat}}$BN grown with naturally abundant boron and nitrogen isotopes: host hBN matrix in either AA' (a) or AB (b) stacking. PL excitation energy = 4.5 eV, sample temperature = 7 K. Hydrostatic pressures (GPa): \{0.04, 0.65, 1.30, 1.97, 2.50, 3.13, 3.74, 4.36, 5.03\} in panel (a), and \{0.46, 1.60, 2.95, 3.48, 4.10, 4.76, 5.42, 5.90, 6.65, 7.15, 7.80, 8.81\} in panel (b), from bottom to top.}
\label{fig7}
\end{figure*}
\end{center}
\begin{center}
\begin{figure}[ht]
\includegraphics[width=0.49\textwidth]{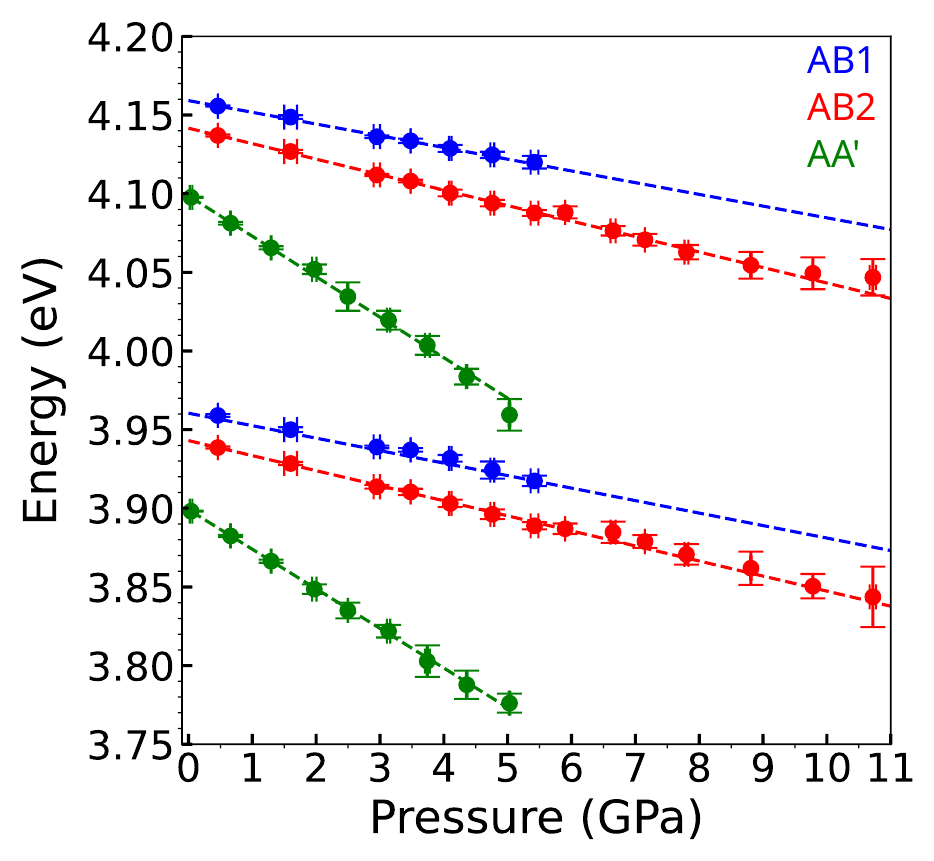}
\caption{Pressure-induced red-shifts of the ZPL and phonon replica at $\sim$3.9 eV for AA' and AB-stacked hBN, with two non-equivalent configurations of the 4 eV-defect in AB-hBN. The slopes of linear fits to the experimental data give the pressure coefficients.}
\label{fig8}
\end{figure}
\end{center}
\begin{center}
\begin{figure}[ht]
\includegraphics[width=0.49\textwidth]{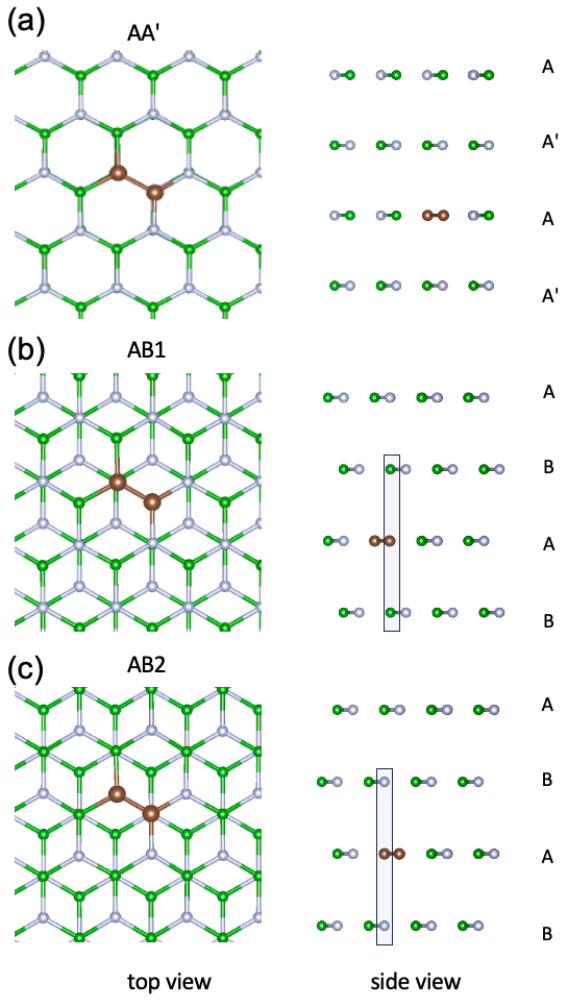}
    \caption{Atomic configurations of the C2 defect in different stacking sequences of hBN: AA' (a),  and AB (b,c) showing the two non-equivalent forms of the carbon dimer in AB-hBN. The so-called AB1 configuration corresponds to C$_{\rm{B}}$C$_{\rm{N}}^{\rm{(B)}}$ (b), while the AB2 one to C$_{\rm{B}}^{\rm{(N)}}$C$_{\rm{N}}$ (c). The non-equivalence is identical for the carbon tetramer and the carbon 6-ring by repetition of the atomic arrangements according to the schemes in Fig.\ref{fig0}. Left column: top-view, along the c-axis. Right column: side-view, perpendicularly to the c-axis. }
\label{fig_3forms}
\end{figure}
\end{center}
\section{Stacking sequence control of the host \MakeLowercase{h}BN matrix}
\label{sec:polytype}
To further screen the different carbon-based candidate structures, we now make use of an original degree of freedom, which is the stacking sequence of the hBN basal planes \cite{gil}. We investigate hBN polytypes either in the standard AA' stacking \cite{pease} or in the AB one, corresponding to the Bernal form of hBN \cite{gil}. By pressure-dependent PL measurements of the 4 eV-defect, we demonstrate that the perturbative response to hydrostatic pressure is stacking-dependent, providing additional key fingerprints for the definitive identification of the 4 eV-defect as a carbon dimer.
\subsection{Samples}
AA'-stacked hBN crystals were synthesized in KSU with the well-established metal flux growth method described in Ref.\cite{pelini}. This method also produces AB-stacked hBN crystals but of reduced size, few tens of micrometers (because of the polytypic nature of the currently available samples \cite{rousseauAB,rousseauG,rousseau,rousseauM}), i.e. a dimension hardly appropriate for high-pressure measurements in a diamond-anvil cell [see Appendix~\ref{sec:PL_Warsaw}] with limited spatial resolution compared to confocal microscopy at ambient pressure \cite{pelini,rousseau}. Although experimental data could be obtained in such samples [see details and Fig.\ref{fig_suppPressure} in Appendix~\ref{sec:C33}], the use of an AB-hBN powder has allowed us to explore the hydrostatic pressure response of AB-stacked hBN [Fig.\ref{fig7}b].

The AB-stacked hBN powder was synthesized by thermochemical conversion of a lithium-modified ammonia borane (NH$_3$BH$_3$) precursor [see Appendix~\ref{sec:Limoges}]. The precursor was placed in a BN crucible in a glove box, then put in a sealed tube under argon atmosphere to avoid oxygen contamination of the samples during the transfer to the furnace. The BN crucible containing the powder was introduced into a silica tube inserted in a horizontal furnace (Thermconcept ROS 50/450/12) under argon flow. The tube was then evacuated to 0.1 mbar and kept under dynamic vacuum for 30 min. Afterwards the silica tube was refilled with purified anhydrous ammonia (99.99\%) to atmospheric pressure. The heating program went as follows: heating to 1000\degreecelsius\, at a speed of 5\degreecelsius/min, dwelling for 2 h and cooling down to room temperature at 5\degreecelsius/min. A constant flow of 120 mL/min passed through the silica tube during the pyrolysis. The flowing gas was switched to N$_2$ after the 2h dwelling at 1000\degreecelsius\, until the end of the treatment. Then an annealing treatment up to 1800\degreecelsius\, was done in a graphite furnace (VHT-GR, Nabertherm) still using a BN crucible. The furnace was put under vacuum (0.1 mbar, 30 min), refilled with nitrogen and maintained under a constant flow of 200 mL/min of N$_2$ during the whole heat treatment. The heating and cooling ramps were of 5\degreecelsius/min and the final temperature was of 1800\degreecelsius\, with a dwelling time of 2 h. The AB-stacking was evidenced by PL spectroscopy in the UV-C and UV-B spectral domains [see Appendix~\ref{sec:SB}, Fig.\ref{fig_suppAB}], where the simultaneous observation of a line at $\sim$6.035 eV and a ZPL doublet for the 4 eV-defect proves the Bernal form of BN \cite{rousseau,rousseauAB}.
\subsection{Experimental results}
Pressure-dependent measurements of the 4 eV-defect emission were performed with the experimental setup described in Appendix~\ref{sec:PL_Warsaw}.

In the case of the standard AA' stacking [Fig.\ref{fig7}a], the emission was studied from ambient pressure up to $\sim$5 GPa. The PL spectrum smoothly red-shifts when raising the hydrostatic pressure, and the ZPL width increases. At 2.5 GPa, the ZPL is too broad to be observed, but the PL emission still peaks at the ZPL energy, with a surrounding spectrum determined by the low-energy acoustic phonons sideband, similarly to temperature-dependent experiments \cite{phuongPRL}. This explains the broad PL bands at high pressures in Fig.\ref{fig7} where the PL spectra, normalized to their maximum intensity, are dominated by the vibronic bands of the color center. From the pressure-dependent PL measurements in Fig.\ref{fig7}a, we extract the pressure-induced red-shifts of the ZPL and $\Delta$-split phonon replica in the AA' stacking [Fig.\ref{fig8}], and we deduce a pressure coefficient of -25.5$\pm$0.5 meV/GPa, in agreement with previous studies \cite{koronski}. 

Pressure-dependent experiments in AB-hBN reveal the striking influence of the stacking sequence on the optoelectronic properties of the 4 eV-defect subjected to hydrostatic pressure. From the PL spectra recorded up to $\sim$9 GPa [Fig.\ref{fig7}b], the pressure-induced perturbation appears to be significantly reduced with respect to AA'-hBN, with a red-shift of only $\sim$100 meV at $\sim$9 GPa in AB-hBN compared to $\sim$125 meV at $\sim$5 GPa in AA'-hBN. Over the full range of investigated pressure, we estimate an average $\sim$3-fold decrease of the pressure-induced red-shift in AB-hBN [Fig.\ref{fig8}].

Because of the staggered stacking of the hBN basal planes in AB-hBN, there are two non-equivalent configurations of the 4 eV-defect, as experimentally demonstrated in Ref.\cite{rousseau}, and schematically represented in Fig.\ref{fig_3forms}. One of the two substitutional carbons has no neighbour in the adjacent planes along the c-axis so that the carbon dimer is either C$_{\rm{B}}$C$_{\rm{N}}^{\rm{(B)}}$ (so-called AB1 configuration, Fig.\ref{fig_3forms}b) or C$_{\rm{B}}^{\rm{(N)}}$C$_{\rm{N}}$ (so-called AB2 configuration, Fig.\ref{fig_3forms}c), in contrast to AA'-hBN where the carbon dimer has only the C$_{\rm{B}}^{\rm{(N)}}$C$_{\rm{N}}^{\rm{(B)}}$ form [Fig.\ref{fig_3forms}a]. This is a singularity of the AB stacking, and we stress that there is only one configuration of the carbon dimer (and by extension of the carbon tetramer and carbon 6-ring) for the AA' but also for the AA and ABC stackings, so that the experimental observation of a ZPL doublet at $\sim$4 eV is a fingerprint of AB-hBN together with a PL line in the UV-C range at $\sim$6.035 eV \cite{rousseau,rousseauAB}. 

This specific ZPL doublet was used for characterizing the AB-hBN powder synthesized by thermoconversion of lithium-modified ammonia borane [see Appendix~\ref{sec:SB}, Fig.\ref{fig_suppAB}]. In the following, the two components of the ZPL doublet at 4.161 and 4.145 eV are labelled AB1 and AB2, respectively. Remarkably, our pressure-dependent PL measurements resolve different pressure-induced red-shifts for the two non-equivalent configurations of the 4 eV-defect. Although the AB1 ZPL energy can not be estimated over the full range of investigated pressure but only until $\sim$5 GPa [see Appendix~\ref{sec:SB}, Fig.\ref{fig_suppFit}], the pressure-induced red-shifts of the AB1 and AB2 ZPLs and their $\Delta$-split phonon replicas are different [Fig.\ref{fig8}], and we estimate values of -7.7$\pm$0.5 and -9.7$\pm$0.3 meV/GPa for AB1 and AB2, respectively. The AB-hBN micro-crystals were similar, with pressure coefficients of -7.4$\pm$0.5 and -10.4$\pm$0.4 that are identical to the data of Fig.\ref{fig8} within experimental error [see Appendix~\ref{sec:C33}, Fig.\ref{fig_suppPressure}].

Although the modification of the stacking sequence of the host hBN matrix only changes the ZPL energy of the 4 eV-defect by $\sim$1 \% from AA' to AB, the huge variations of the perturbative response to hydrostatic pressure in the two hBN polytypes reveal the importance of considering the interlayer coupling. Together with the more subtle difference in the pressure-induced red-shifts of the two non-equivalent configurations of the color center in AB-hBN, these experimental results provide crucial inputs for further elucidating the defect structure.
\subsection{\textit{Ab initio} calculations}
\begin{center}
\begin{figure}[ht]
\includegraphics[width=0.49\textwidth]{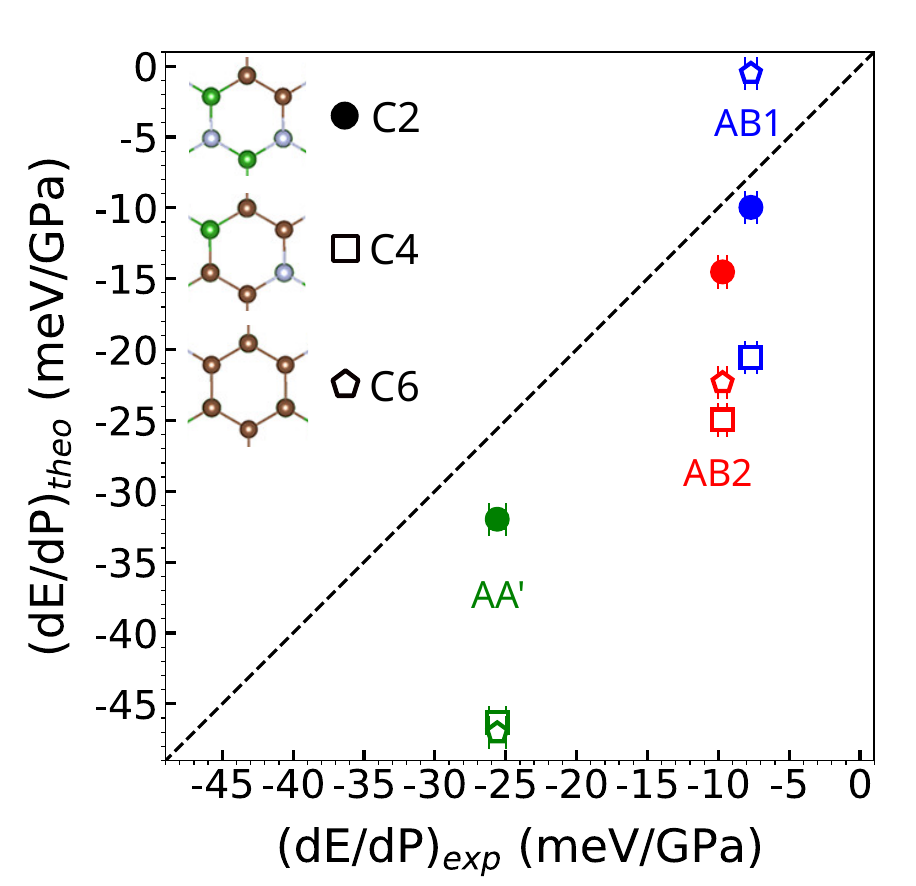}
\caption{Pressure coefficients of the ZPL for the three configurations of the carbon-based UV-color center in AA' and AB-stacked hBN: experimental data on the x-axis vs theoretical values on the y-axis for the C2 (full circles), C4 (empty squares) and C6 (empty pentagons) structures.}
\label{fig9}
\end{figure}
\end{center}

We investigated the impact of hydrostatic pressure on the PL spectra within our theoretical framework [see Appendix~\ref{sec:DFT_addit}]. The application of hydrostatic pressure results in a compressive strain, with the majority of the strain being absorbed along the c-direction ($\sim$12\% compression in both AA’ and AB stacking sequences at 10 GPa), while the basal plane lattice constants experience smaller variations ($\sim$0.7\% compression). Fig.\ref{fig9} illustrates the comparison of the pressure coefficients for the various carbon-based defects. Our calculations reveal that the values obtained for the carbon dimer (C2) closely match the experimental data, reaching again nearly quantitative agreement. For the carbon tetramer (C4), the relative variations of the computed pressure-induced shifts from one defect configuration to another are in agreement with our experimental findings, but the absolute values of the pressure coefficients are significantly overestimated by a factor of $\sim$2 with respect to our measurements. The discrepancy with the experimental data is even more pronounced for the carbon 6-ring (C6), for which the AB1 ZPL is calculated to be almost insensitive to hydrostatic pressure.

To elucidate the difference of pressure effects on the C2 defect in AA’ and AB stacking sequences, we have analysed its electronic properties, shown in Tab. \ref{tab_ts1_0} [Appendix~\ref{sec:SI_DFT}]. The primary effect of hydrostatic pressure is the compression of the interlayer spacing, leading to an increased screening effect. Consequently, this results in the stabilization of defect levels with respect to the band edges. In the case of the AA’ sequence, both C$_{\rm{B}}$ and C$_{\rm{N}}$ are directly linked to nitrogen and boron atoms from adjacent layers, respectively [Fig.\ref{fig_3forms}a]. Therefore, both occupied and empty levels are stabilized, resulting in a red-shift of the PL signal. This effect can also be interpreted as a suppression of the exchange interaction between the donor and acceptor defects due to enhanced screening. Conversely, in the AB stacking configuration, only one carbon atom of the pair (C$_{\rm{N}}$ in AB1, and C$_{\rm{B}}$ in AB2) interacts directly with neighboring layers. Consequently, the empty defect level in AB1 and an occupied level in AB2 are primarily stabilized, while their counterparts are less affected. As a consequence, the red-shift of the PL signal is less pronounced compared to the AA’ sequence. This elucidates the difference in pressure effects between AA' and AB stacking arrangements in terms of electronic response.

To further understand the difference between AB1 and AB2 configurations of the C2 defect, we also considered the effects of structural relaxation, with relevant data summarised  in Tab.~\ref{tab_ts1} [Appendix~\ref{sec:SI_DFT}]. Our analysis, consistent with our previous research \cite{li2024exceptionally}, reveals that interlayer polarization occurs as electron density leaks from the lone pairs of nitrogen to boron atoms. In the AB2 configuration [Fig.\ref{fig_3forms}c], where the donor carbon occupies the boron site (C$_{\rm{B}}$), a strong repulsive interaction with neighboring nitrogen atoms from adjacent layers is observed in the excited state upon a back-electron transfer. This interaction causes the defect to slide along the symmetry axis; the degree of sliding increases under pressure, resulting in a decrease in the N-C-N angle. Consequently, this distortion leads to a large increase in the total Huang-Rhys factor for this configuration, which is also reflected in our experimental spectra. In contrast, no strong interaction between C$_{\rm{N}}$$^+$ and adjacent boron atoms is observed, as they lack electron density. Therefore, the AB1 and AB2 configurations of the C2 defect show a different structural response to hydrostatic pressure, as seen in our experiments.

   
\section{Conclusion and outlook}
We have presented a general methodology for point defects identification, combining isotope substitution and polytype control, with a systematic comparison between experiments and first-principles calculations. We have applied it to the 4 eV-defect in hBN, and tackled the long-standing issue of the nature of this ubiquitous color center in hBN. By implementing isotopic purification of the host hBN matrix, followed by isotope-selective carbon doping of hBN, we have reduced the number of candidate structures to a few carbon-based centers. Then, by playing with the stacking sequence of the host matrix in two hBN polytypes, we have revealed different perturbative responses to hydrostatic pressure for the non-equivalent configurations of the 4 eV-defect in AA' and AB-stacked hBN. Our comprehensive set of experimental and theoretical results demonstrate that the 4 eV-defect is a carbon dimer in the hBN honeycomb lattice.

Our study is an important milestone in hBN research, where the impact of carbon is the subject of an intense debate. The demonstration of the 4 eV-defect to being a carbon dimer sets the basis for a better understanding of the incorporation of carbon in hBN and its possible role in hBN dopings. The carbon dimer is an isoelectronic defect in the hBN honeycomb lattice, but its dissociation may lead to the formation of donor-acceptor pairs, signatures of which have been reported in the optical response of hBN \cite{museur,du,bourrellier,vokhmintsev2,korona}. Both n- and p-dopings are still in their infancy in hBN \cite{jiang,mballo,lu}, and our results may shed a new light for elucidating the striking persistent photo-conductivity phenomenon reported in hBN grown by metal-organic chemical vapor deposition \cite{perepeliuc}. Moreover, the identification of the carbon dimer raises the question of the optical responses of the carbon tetramer and carbon 6-ring, which are also expected in the UV-B spectral range. Further work appears necessary to determine the growth conditions that can lead to suitable concentrations of these carbon-based centers for optical experiments. Because carbon 6-rings are predicted to have a large binding energy favoring the formation of carbon clusters \cite{li}, tracking this specific defect may be crucial for understanding the synthesis of hBN polytypes, in particular the Bernal form appearing during the fabrication of hBN crystals under the addition of a large amount of carbon \cite{rousseauAB,rousseauG,rousseau,rousseauM}. Finally, in the context of quantum technologies, our method may stimulate follow-up studies in order to identify the carbon-based defects having an optical response in the visible and near-infrared domains, and acting as single photon emitters \cite{mendelsohn} and spin defects \cite{stern,chejanovsky}.

We found in hBN in this study and in a previous work \cite{li2024exceptionally} that a strong coupling appears between the hBN layers because of the Coulombic interaction between the positively polarized boron atoms and negatively polarized nitrogen atoms in neighbor hBN sheets. This is clearly manifested in the PL signals of planar defects such as the 4.1-eV emitter.
Beyond hBN, our work opens novel perspectives for identifying point defects in materials where the stacking order could be changed, such as SiC \cite{awschalom}, thus providing a systematic method combining polytype control and isotope substitution, for both the host matrix and the impurities. It can be expected that the PL signals can be tuned by polytype and isotope control of the respective layers when the host material contains polarized bonds which indeed appear for hetero-atomic 2D materials. Furthermore, our results may guide future theoretical and experimental studies on heterostructures of 2D materials to engineer the polarization of bonds in subsequent sheets that can tune the optical signals of planar defects in the target 2D layer.

\textbf{Acknowledgements}

We gratefully acknowledge C. L'Henoret and T. Cohen for their technical support at the mechanics workshop. This work was financially supported by the BONASPES project (ANR-19-CE30-0007), the ZEOLIGHT project (ANR-19-CE08-0016), and the HETERO-BNC project (ANR-20-CE09-0014-02). Support for hBN crystal growth at KSU was provided by the Office of Naval Research, Award no. N00014-22-1-2582. Support for hBN crystal growth at SEU was provided by the National Science Foundation of China, Grant No. 11674053. A.G.\ acknowledges the support from the National Research, Development and Innovation Office of Hungary (NKFIH) in Hungary for the Quantum Information National Laboratory (Grant No.\ 2022-2.1.1-NL-2022-0000) and the EU HE project SPINUS (Grant No.\ 101135699). A part of the calculations was performed using the KIF\"U high-performance computation units. A.P.\ acknowledges the financial support of Janos Bolyai Research Fellowship of the Hungarian Academy of Sciences.

$^\ast$e-mail: gali.adam@wigner.hun-ren.hu, guillaume.cassabois@umontpellier.fr
\begin{appendix}
\section{METHODS}
    \subsection{hBN growth by the metal flux method}
    \label{sec:BN_SEU}
Samples from SouthEast University (SEU series) were grown by using the metal flux method. For boron isotope controlled single crystal growth, high purity iron granules (Fe, 5N purity) and chromium granules (Cr, 4N purity) were used as metal flux. Powders of $^{10}$B (85 at\%) or $^{11}$B (99 at\%) were used as source materials for the synthesis of $^{10}$B-enriched hBN or h$^{11}$B$^{14}$N crystals, respectively. For h$^{\rm{Nat}}$BN crystal growth, iron (Fe, 5N purity), nickel (Ni, 6N purity), chromium (Cr, 4N purity) and cobalt (Co, 5N purity) were used as metal flux, with high-purity h$^{\rm{Nat}}$BN powder as source materials. Carbon powder (with naturally abundant isotopes, i.e. 99\% $^{12}$C and 1\% $^{13}$C) was included in all growths to promote the hBN single crystal formation \cite{wan}. Typically, the metals, boron source ($^{10}$B, $^{11}$B or hBN powders) and carbon powder were weighted in ratios of 100:10:2, mixed together and then loaded into an alumina crucible, which were then placed inside an alumina tube furnace. The furnace was purged 3-5 times by nitrogen gas (purity $\sim$99.9\%, naturally abundant isotopes) after being evacuated to a vacuum of $\sim$5 Pa prior to the growth in order to minimize residue gas in the furnace tube. The temperature was then ramped to 1550\degreecelsius\, at rate of 6\degreecelsius/min, held at 1550\degreecelsius\, for 12$\sim$24 hours, then cooled slowly to 1400\degreecelsius\, at a rate of 4\degreecelsius/h to obtain high-quality hBN crystals. Finally, the sample was cooled down to room temperature naturally. During all these processes, the flow rate of N$_2$ gas was kept at $\sim$250 sccm.
    \subsection{Photoluminescence spectroscopy}
    \subsubsection{Ambient pressure measurements}
    \label{sec:PL_MPL}
The emission spectrum of the 4 eV-defect was recorded by PL spectroscopy with two experimental setups devoted to macro-PL \cite{eliasNatCom} and micro-PL \cite{pelini}, with spatial resolutions of 50 $\micro$m and 300 nm, respectively. The PL was excited above the hBN bandgap by the fourth harmonic of a cw mode-locked Ti:Sa oscillator, tunable from 193 nm to 205 nm with trains of 140 fs-pulses at 80 MHz repetition rate. The samples were held on the cold finger of closed-cycle cryostats for temperature-dependent measurements from 10 K to room temperature. Samples of a given series were mounted side-by-side in the cryostat in order to perform the PL experiments in the exact same conditions and avoid calibration issues, which could be detrimental for resolving the faint isotopic shifts of the PL lines. The detection system was composed of a f = 500 mm Czerny–Turner monochromator, equipped with a 1800 grooves mm$^{-1}$ grating blazed at 250 nm, and with a backilluminated CCD camera (Andor Newton 920), with a quantum efficiency of 50\% at 210 nm.

The PL spectra of the SEU series (Section \ref{sec:isotopehBN}) and of the $^{\rm{Nat}}$C:hBN and $^{13}$C:hBN crystals (Section \ref{sec:isotopeC}) were acquired by macro-PL experiments, and the KSU series (Section \ref{sec:isotopehBN}) was studied by micro-PL.
    \subsubsection{Pressure-dependent measurements}
    \label{sec:PL_Warsaw}
Low-temperature emission spectra under continuous-wave excitation were obtained using the 275.4 nm-line of a Coherent Innova 400 Ar-ion laser. The sample was mounted inside a closed-cycle helium cryostat. The spectra were dispersed by a Horiba Jobin-Yvon FHR 1000 spectrometer, and the signal was detected by a liquid nitrogen-cooled CCD camera.

High hydrostatic pressure was obtained by using a low-temperature diamond anvil cell (CryoDAC LT, easyLab Technologies Ltd). Argon was used as a pressure-transmitting medium. The diamond anvil cell was mounted in an Oxford Optistat CF cryostat equipped with a temperature controller for low-temperature measurements. Samples with a diameter of $\sim$100 $\mu$m and a thickness of $\sim$30 $\mu$m were loaded into the cell along with a small ruby ball. The R1-line ruby luminescence, excited by the second harmonic of an YAG:Nd laser (532 nm), was used for pressure calibration. The half-width of the R1 line was also used for monitoring hydrostatic conditions in the diamond anvil cell.
    \subsection{\textit{Ab initio} calculations}
    \label{sec:DFT_addit}
The first-principles calculations were conducted using the Vienna \textit{ab initio} simulation package (VASP) code~\cite{kresse1996efficiency, kresse1996efficient} based on projector augmented wave (PAW) potentials~\cite{blochl1994projector, kresse1999ultrasoft}. The cutoff energy for the plane-wave expansion of the wavefunction was set to 450~eV. To prevent defect-defect interactions, a $9\times9$ bilayer hBN supercell model was utilised with single $\Gamma$-point sampling. The interlayer van der Waals (vdW) interaction was accounted for using the DFT-D3 method developed by Grimme~\cite{grimme2010consistent}.

For better comparison with experimental data, the lattice was optimized using the generalized gradient approximation of Perdew, Burke, and Ernzerhof (PBE)~\cite{perdew1996generalized} followed by electronic property calculations using the hybrid density functional of Heyd, Scuseria, and Ernzerhof (HSE)~\cite{heyd2003hybrid}. The exchange contribution was modified to $\alpha = 0.32$ to reproduce the experimental optical gap around 6 eV. Excited state calculations were performed using the $\Delta$SCF method~\cite{gali2009theory}. We applied a {von Barth} correction for the total energy of the excited state~\cite{mackoit}. A convergence threshold of 0.01~eV\AA$^{-1}$ was set in the force calculations. The  spectral simulations were performed using the Huang-Rhys theory \cite{huang1950theory} with the details, provided in Ref.~\cite{li2024exceptionally}.

The effects of isotope composition on zero-phonon line and sideband were considered by adjusting the phonon modes. To alleviate the computational burden of phonon calculations for different compositions, the following procedure was implemented. Initially, (squares of) phonon frequencies ($F$) and dynamic matrix ($DM$) were obtained for defects in hBN of natural abundance. The mass-scaled Hessian matrix ($H$) was then computed as $H=DM\times F \times{DM}^{-1}$, and the elements were rescaled with the masses of interest to calculate new phonon modes and frequencies for the specific composition. This approach yielded numerically equivalent photoluminescence spectra as direct calculations.
    \subsection{Synthesis of lithium-modified ammonia borane}
    \label{sec:Limoges}
All chemical products were stored and handled in an argon-filled glove box (Jacomex, Campus-type; O$_2$ and H$_2$O concentrations kept at $\leq$ 0.1 ppm and $\leq$ 0.8 ppm, respectively). The cleaned ball milling reactor and crucibles were stored in an oven at 100\degreecelsius\, overnight before being introduced in the glove box to be used. Lithium amide, LiNH$_2$ (95\%), and ammonia borane complex, NH$_3$BH$_3$ (97\%) were purchased from Sigma-Aldrich and used as received.

Li-modified ammonia borane was synthesized by solid state reaction. For a fixed molar ratio $n_{\rm{NH_3BH_3}}/n_{\rm{Li}}$ of 3, approximately 802 mg of NH$_3$BH$_3$ and 198 mg of LiNH$_2$ were placed in a stainless-steel ball-milling reactor, in the glove box. 6 stainless-steel beads of 10 mm in diameter were placed on top of the powders before closing the reactor. The products were then ball-milled on a pulverisette 7 premium line apparatus from Fritsch with the following parameters: 5 cycles of 1 min-runs followed by 2 min pause (very slow rotation) at 250 revolutions per minute (rpm), reverse mode enabled and venting between each cycle. Evacuation of excess pressure coming from the reaction between LiNH$_2$ and NH$_3$BH$_3$ (mainly NH$_3$ and H$_2$) was performed via a valve of the ball-milling reactor. This reaction leads to the formation of a very viscous white and opaque liquid which solidified after 2 or 3 days. The ball-milling reactor was left open in the glove box until the white powder formed and could be easily retrieved.
\section{ADDITIONAL INFORMATION}
    \subsection{Photoluminescence experiments}
        \subsubsection{AB-hBN powder}
            \label{sec:SB}
\begin{center}
\begin{figure}[ht]
\includegraphics[width=0.49\textwidth]{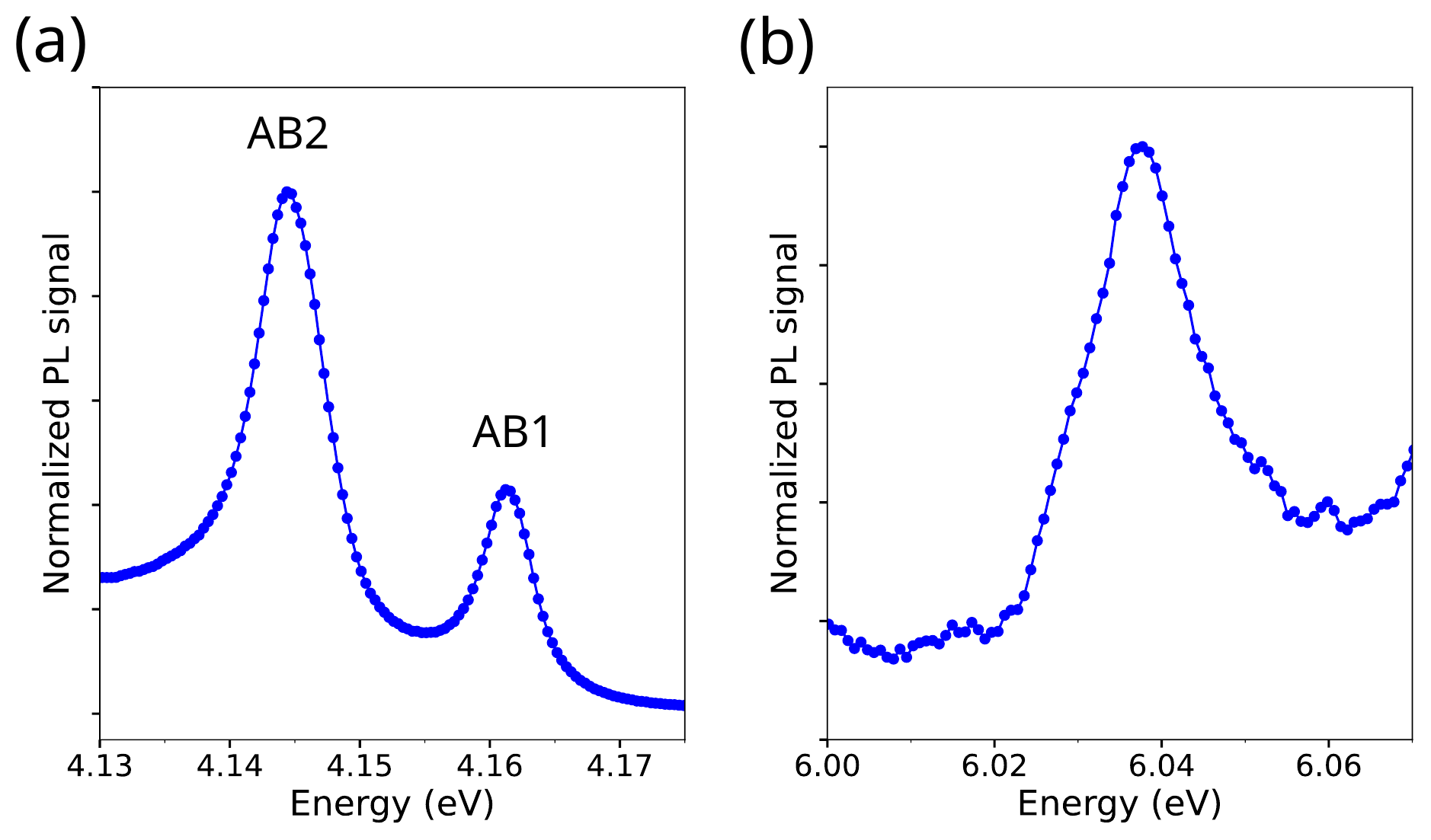}
\caption{Photoluminescence spectra of the AB-hBN powder synthesized by thermoconversion of a lithium-modified ammonia borane precursor: detection in the UV-B (a) and UV-C (b) spectral ranges. Ambient pressure, sample temperature = 8 K, PL excitation energy = 6.32 eV.}
\label{fig_suppAB}
\end{figure}
\end{center}
\begin{center}
\begin{figure}[h]
\includegraphics[width=0.49\textwidth]{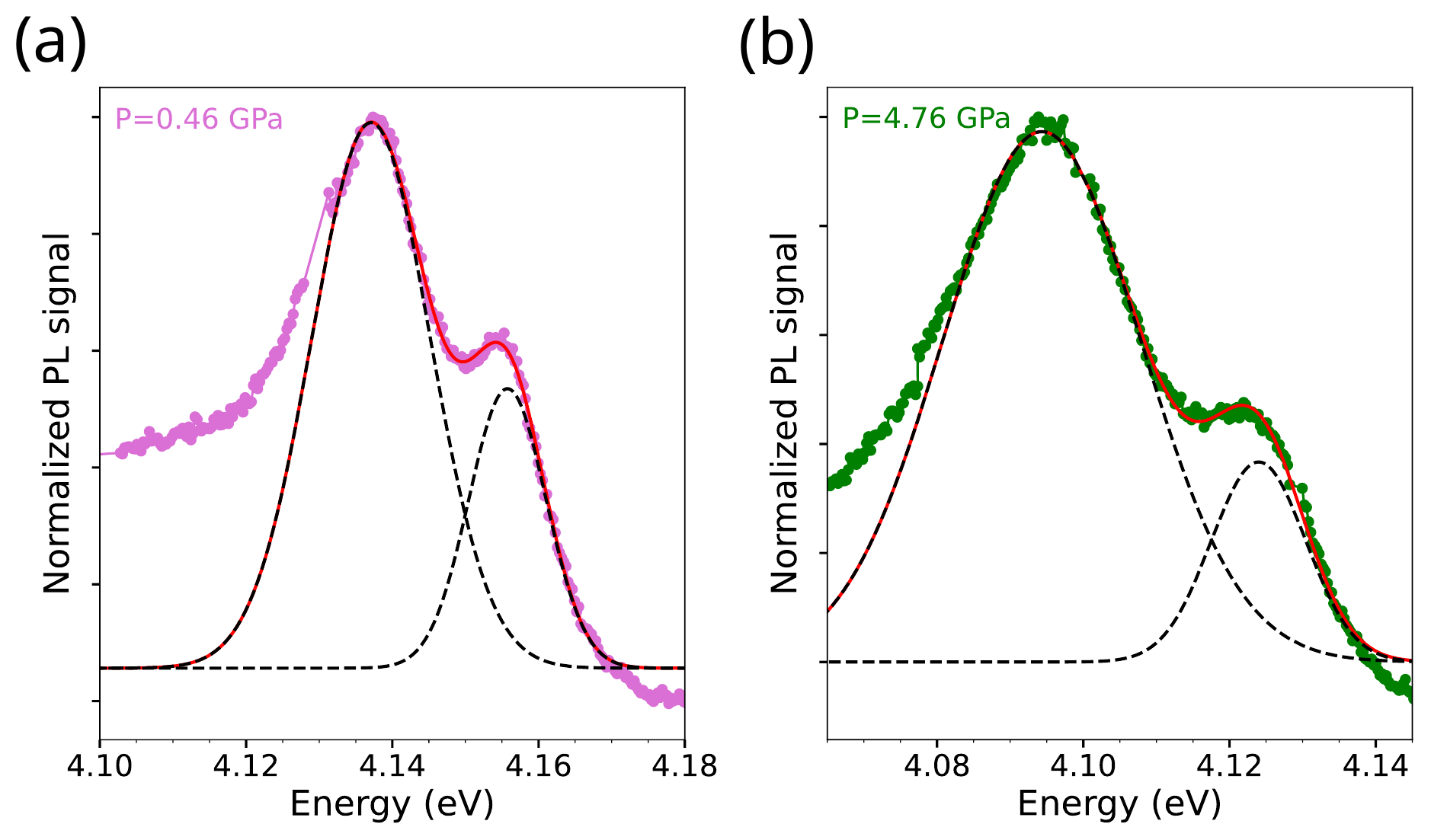}
\caption{Photoluminescence spectra of the AB-hBN powder at hydrostatic pressures of 0.46 GPa (a) and 4.76 GPa (b): experimental data (symbols), two Gaussian-fit (solid line).}
\label{fig_suppFit}
\end{figure}
\end{center}

The AB-stacking of hBN synthesized from a lithium-modified ammonia borane precursor was evidenced by PL spectroscopy in the UV-C and UV-B spectral domains [Fig.\ref{fig_suppAB}], where the simultaneous observation of a line at $\sim$6.035 eV and a ZPL doublet for the 4 eV-defect proves the Bernal form of BN \cite{rousseau,rousseauAB}. In Ref.\cite{rousseauAB}, spatially-resolved experiments in AA'- and AB-stacked hBN crystals demonstrated a one-to-one correlation between inversion symmetry breaking probed by second harmonic generation and the detection of an intense PL line at $\sim$6.035 eV, i.e. the specific signature of the noncentrosymmetric Bernal stacking.

The pressure-dependent energies of the two ZPLs of the 4 eV-defect in AB-stacked hBN were estimated by a two-Gaussian fit [Fig.\ref{fig_suppFit}], allowing a quantitative analysis of the AB1 component (high-energy line at $\sim$4.161 eV for ambient pressure) at hydrostatic pressures up to 4.76 GPa.

        \subsubsection{AB-hBN micro-crystals}
            \label{sec:C33}
\begin{center}
\begin{figure}[h]
\includegraphics[width=0.49\textwidth]{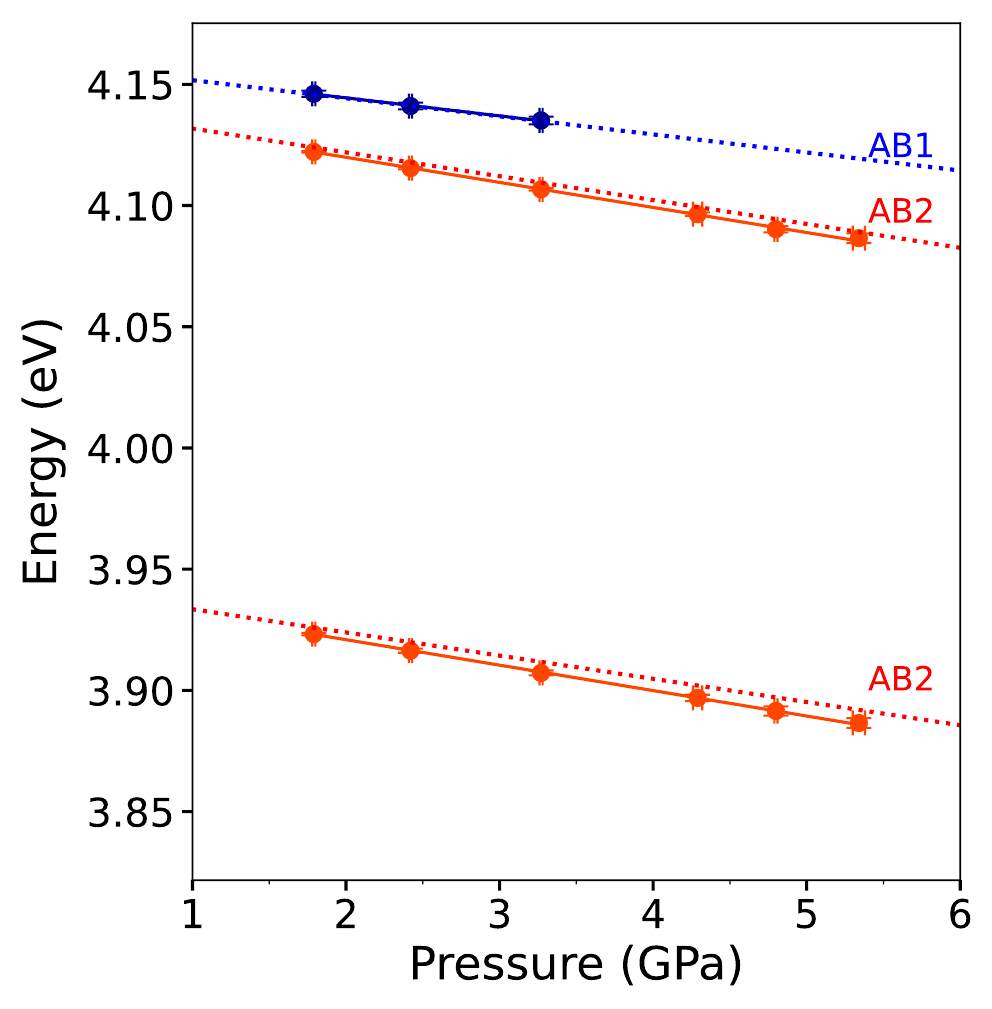}
\caption{Pressure-induced red-shifts of the ZPL doublet and phonon replica of the AB2-ZPL in AB-stacked hBN micro-crystals: experimental data (symbols), replicated fit to the data in the AB-hBN powder from Fig.\ref{fig8} (dotted lines).}
\label{fig_suppPressure}
\end{figure}
\end{center}

The polytypic nature of the currently available hBN samples synthesized by the metal flux growth method \cite{rousseau,rousseauAB} reduces the size of AB-stacked hBN crystals to few tens of micrometers, i.e. a dimension hardly appropriate for high-pressure measurements in a diamond-anvil cell, with limited spatial resolution compared to confocal microscopy at ambient pressure \cite{pelini,rousseau}. Nevertheless, experiments could be performed in a few samples, leading to results identical to the measurements performed in the AB-stacked hBN powder [dotted lines in Fig.\ref{fig_suppPressure}] within experimental error.
\subsection{\textit{Ab initio} calculations}
\label{sec:SI_DFT}
\begin{center}
\begin{figure}[ht]
\includegraphics[width=0.49\textwidth]{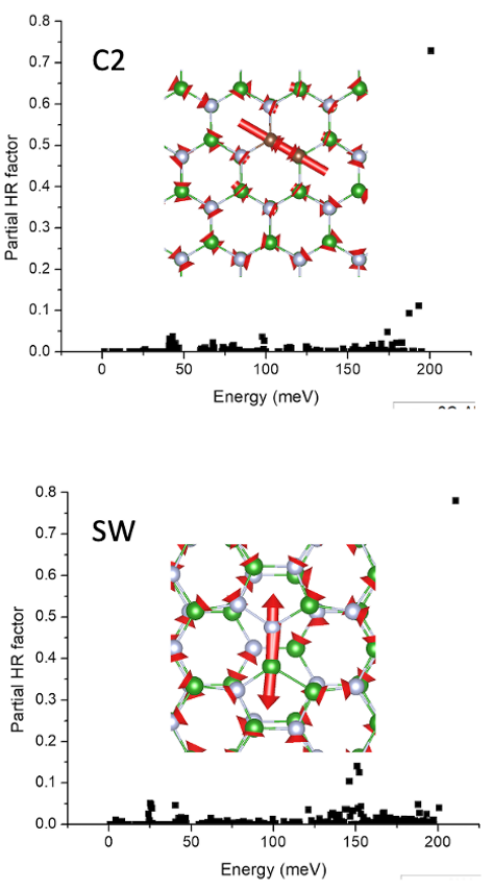}
\caption{Partial Huang-Rhys (HR) factors for each vibration mode, calculated for the C2 and SW defects in the AA' stacking sequences of hBN. The red arrows show the atomic contributions to the leading mode at $\sim$200 meV.}
\label{fig_ts1}
\end{figure}
\end{center}

\begin{center}
\begin{figure}[ht]
\includegraphics[width=0.49\textwidth]{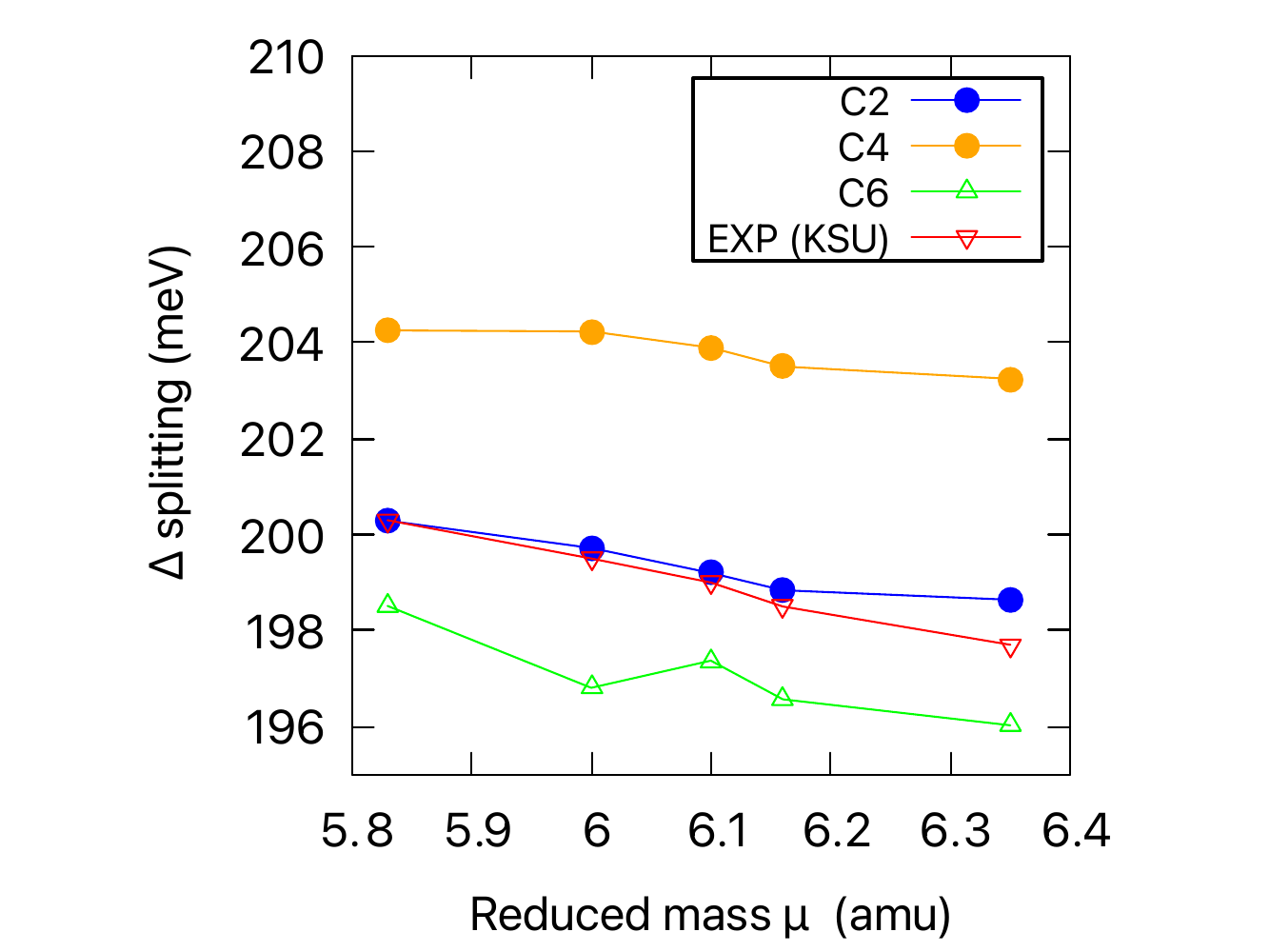}
\caption{Calculated isotopic shifts of the energy splitting $\Delta$ between the zero-phonon line at $\sim$4.1 eV and the phonon replica (as defined in Fig.\ref{fig1}) for the carbon defects, compared with the experimental values.}
\label{fig_ts2}
\end{figure}
\end{center}

\begin{center}
\begin{figure}[ht]
\includegraphics[width=0.49\textwidth]{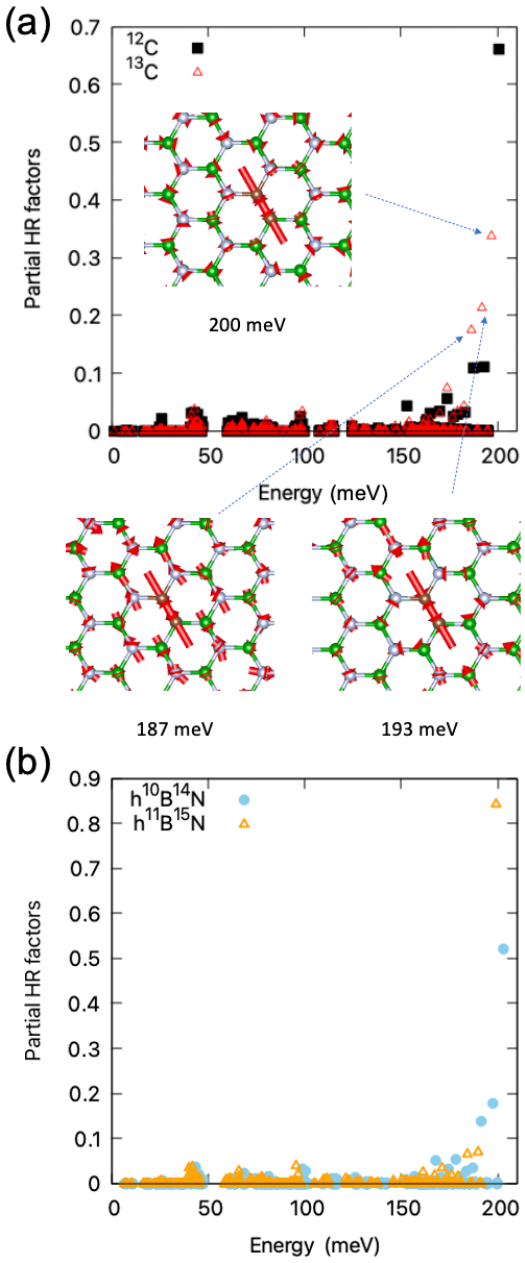}
\caption{Partial Huang-Rhys (HR) factors for each vibration mode, calculated (a) for the $^{12}$C2 and $^{13}$C2 defects in the AA' stacking sequences of hBN and (b) for the $^{12}$C2 defect in h$^{10}$B$^{14}$N and h$^{11}$B$^{15}$N. The red arrows show the atomic contributions to the three leading modes, contributing to the phonon replica.}
\label{fig_ts1a}
\end{figure}
\end{center}

\begin{table}
\begin{ruledtabular}
\begin{tabular}{l|ccc}
         & AA'    & AB1    & AB2      \\
         \hline
0 GPa    &        &        &          \\
CBM      & 6.09 & 6.11 & 6.12   \\
empty    & 5.62 & 5.55 & 5.65   \\
occupied & 0.72 & 0.59 & 0.70   \\
$\Delta$E & 4.9 &  4.96 & 4.95   \\
\hline
10 GPa   &        &        &          \\
CBM      & 5.99 & 5.97  & 5.98	   \\
empty    & 5.49 & 5.45  &  5.70   \\
occupied & 0.75 & 0.57 &  0.81  \\
$\Delta$E & 4.74 &  4.88 & 4.89  \\
\hline
$\Delta$$\Delta$E & 0.16 &  0.08 & 0.06
\end{tabular}
\end{ruledtabular}
\caption{Electronic properties of C2 defect in the ground state in different stacking sequences at 0 and 10 GPa. The energy levels given in eV are aligned to the valence band maximum of each configuration. CBM is the conduction band minimum. $\Delta$E is the difference between the energies of occupied and empty defect levels.  $\Delta$$\Delta$E is an electronic contribution to the pressure coefficient.}
\label{tab_ts1_0}
\end{table}

\begin{table}
\begin{ruledtabular}
\begin{tabular}{l|ccc}
0 GPa   & $\text{AA}^{\prime}$  & AB1  & AB2 \\
\hline
C-C bond gs & 1.367 & 1.371 & 1.360 \\
N-C-N gs& 179.78 & - & 179.75 \\
B-C-B gs& 177.33 & 177.32 & - \\
C-C bond ex& 1.472 & 1.470 & 1.461 \\
N-C-N ex& 177.33 & - & 177.08 \\
B-C-B ex& 178.04 & 178.02 & - \\
HR factor& 2.09 & 2.15 & 2.31 \\
\hline
10 GPa   & $\text{AA}^{\prime}$  & AB1  & AB2 \\
\hline
C-C bond gs& 1.356 & 1.364 & 1.349 \\
N-C-N gs& 179.23 & - & 179.20 \\
B-C-B gs& 177.33 & 177.17 & - \\
C-C bond ex& 1.458 & 1.452 & 1.451 \\
N-C-N ex& 176.08 & - & 174.77 \\
B-C-B ex& 176.70 & 176.90 & - \\
HR factor& 2.53 & 2.11 & 3.60 \\
\end{tabular}
\end{ruledtabular}
\caption{Relevant structural parameters computed for the C2 defect in the ground (gs) and excited (ex) states in different stacking sequences at 0 and 10 GPa. The angles are defined between the respective nearest atoms in the adjoined layers. HR is for Hyang-Rhys factors.}
\label{tab_ts1}
\end{table}

\begin{center}
\begin{figure*}[ht]
\includegraphics[width=0.99\textwidth]{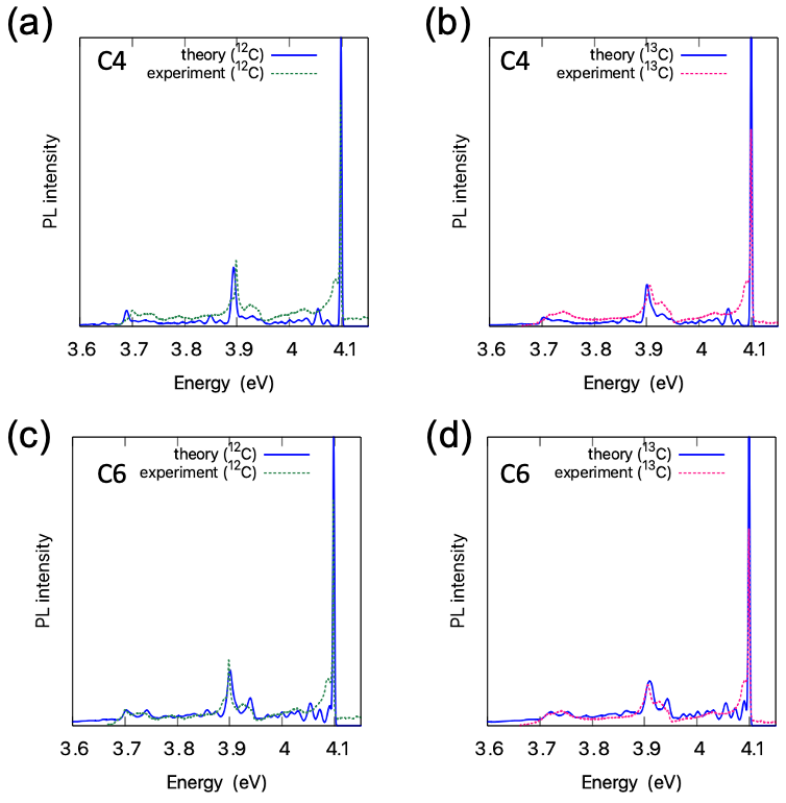}
\caption{Simulated impact of the isotopic substitution ($^{12}$C vs $^{13}$C) on the PL spectra of the C4 and C6 defects. The experimental results are from Fig.\ref{fig6}.}
\label{fig_ts3}
\end{figure*}
\end{center}

\end{appendix}

\end{document}